\RequirePackage{fix-cm}
\documentclass[twocolumn,epjc3]{svjour3}  
\smartqed  
\RequirePackage{graphicx}
\usepackage{subfigure}

\journalname{Eur. Phys. J. C}
\begin{document}

\title{Revisiting Chaplygin gas cosmologies with the recent observations of high-redshfit quasars
}


\author{Jie Zheng\thanksref{addr1}
        \and
        Shuo Cao\thanksref{e1,addr1} 
        \and
        Yujie Lian\thanksref{addr1}
        \and
        Tonghua Liu\thanksref{addr2}
        \and
        Yuting Liu\thanksref{addr1}
        \and
        Zong-Hong Zhu\thanksref{e2,addr1,addr3}
}

\thankstext{e1}{e-mail: caoshuo@bnu.edu.cn}
\thankstext{e2}{e-mail: zhuzh@bnu.edu.cn}

\institute{Department of Astronomy, Beijing Normal University, Beijing 100875, China \label{addr1}
           \and
           School of Physics and Optoelectronic, Yangtze University, Jingzhou 434023, China \label{addr2}
           \and
           School of Physics and Technology, Wuhan University, Wuhan 430072, China \label{addr3}
}

\date{Received: date / Accepted: date}

\maketitle

\begin{abstract}
In this paper, we use the latest observations of quasars covering the redshift range of $0.04<z<5.1$ to investigate a series of Chaplygin gas models as candidates for unified dark matter and dark energy. Based on different combinations of available standard candle and standard ruler data, we put constraints on the generalized Chaplygin gas (GCG), modified Chaplygin gas (MCG), new generalized Chaplygin gas (NGCG) and viscous generalized Chaplygin gas (VGCG) models. Moreover, we apply Jensen-Shannon divergence (JSD), statefinder diagnostics, and the deviance information criterion (DIC) to distinguish these CG models, based on the statistical results derived from Markov chain Monte Carlo method. The results show that (1) The standard ruler data could provide more stringent constraints on the cosmological parameters of different CG models considered in this analysis. Interestingly, the matter density parameter $\Omega_{m}$ and Hubble constant $H_{0}$ derived from the available data are well consistent with those from the Planck 2018 results; (2) Based on the statistical criteria JSD, our findings demonstrate the well consistency between Chaplygin gas and the concordance $\Lambda$CDM model. However, in the framework of statefinder diagnostics, the GCG and NGCG models cannot be distinguished from $\Lambda$CDM, while MCG and VGCG models show significant deviation from $\Lambda$CDM in the present epoch; (3) According to the the statistical criteria DIC, we show that the MCG and VGCG models have substantial observational support from high-redshfit quasars, whereas the GCG and NGCG models miss out on the less observational support category but can not be ruled out.  \\

\end{abstract}

\section{Introduction}
\label{intro}
The analysis of various observational data, including Type Ia supernovae (SNe Ia) \cite{SNe1,SNe2}, baryon acoustic oscillation (BAO) \cite{eisenstein20005}, and cosmic microwave background (CMB) \cite{CMB} suggest that the present universe is undergoing an accelerated phase of expansion \cite{acc_universe}. Different suggestions have been put forward to understand this phenomenon, with the inclusion of exotic dark energy (DE) with negative pressure on the right-hand side of the Einstein equation. The earliest and simplest model for DE is the cosmological standard $\Lambda$CDM model, which is in good agreement with recent observations but embarrassed by the well known coincidence problem and the fine-tuning problem \cite{weinberg1,weinberg2}. Meanwhile, the existence of dark matter (DM), which constitutes the major component of the matter density in our Universe, is the other primary indicator for the limitation of our knowledge of physics laws \cite{Cao2021,2022A&A...659L...5C}. In recent times, scholars proposed that a fluid called Chaplygin gas could provide a possible solution to unify two uncharted territories, mimicing the effects of DM in the early times and DE in the late times \cite{kamenshchik2001}. Specially, the Chaplygin gas obeys the exotic equations of state:
\begin{equation}
p_{}=-\frac{A}{\rho},
\label{eq:CG}
\end{equation}
where $p$ and $\rho$ denote the pressure and energy density, respectively. $A$ is a positive constant. Unlike quintessence, which describes the transition from the quasi-exponential expansion of the early universe to a power law expansion to explain the present acceleration of the universe but fails to avoid fine-tuning in explaining the cosmic coincidence problem, the Chaplygin gas (CG) model provided an alternative way to account for the accelerating universe by describing a transition from an epoch filled with dust-like matter to an accelerating universe. Additionally, they predicted that the cosmological constant was variable. In particular, the Chaplygin gas behaves as a pressureless fluid at higher redshifts and as a cosmological constant at lower redshifts, which tends to promote expansion. In addition, the equation of state of CG shows a well-defined connection with string and brane theories \cite{kamenshchik2001,Bento2003}. 
However, several fatal drawbacks appeared in CG models. There is unexpected blowup in the DM power spectrum \cite{AH32,AH33} in the framework of the CG model, and the CG model is in disagreement with the observations, such as Type Ia supernovae \cite{GCG_SNe1,GCG_SNe2,GCG_SNe3}, X-ray gas mass fraction of clusters \cite{zhu2004}, Hubble parameter-redshift data\cite{GCG_hz} and gamma-ray bursts \cite{GCG_Gamma}. 
Therefore, generalized Chaplygin gas (GCG) model was proposed \cite{Amendola2003,Bento2003}, which is capable of explaining the background dynamics of the early and late universe and is in good agreement with recent observations. The effective equation of state of GCG, given by $p=\alpha\rho$, proves the evolution of a universe evolving from a phase dominated by non-relativistic matter to a phase dominated by a cosmological constant through an intermediate period. 
There are some undesirable features of the GCG power spectrum caused by adiabatic pressure perturbation, which is produced from a nonzero $\alpha$ \cite{Amendola2003,Thakur2019}. As a result, \cite{MCG2002,MCG2004} proposed the ``modified" Chaplygin gas (MCG) model, which considered an interpolation between standard fluids at high energy densities and Chaplygin gas fluids at lower energy densities.
Another generalization is dubbed new generalized Chaplygin gas model (NGCG), which was proposed by \cite{NGCG2006}. Since the equation of state of dark energy still cannot be determined exactly, they argued that the GCG model could be accommodated to any possible X-type dark energy with constant $\omega$, dual to an interacting XCDM parametrization scenario. In the framework of the NGCG model, it is not only described by Chaplygin gas fluid but also exhibits dust-like matter in the early universe and X-type dark energy in the late universe. Up to now, the nature of dark energy and 
dark matter is still unknown. It is reasonable to consider other forms of dark energy models or further generalize the GCG model. For instance, \cite{VGCG2006} considered a phenomenological model that consists of viscous effects and the features of GCG, dubbed viscous generalized Chaplygin gas (VGCG), which is able to eliminate the problems raised by only dissipative fluids and explain the dynamics of the universe.

With so many GG cosmologies proposed in the literature, it is rewarding to determine which model is strongly supported by the currently available astrophysical probes. There are two general types of distance indicators at present: standard candles (SNe Ia and quasars), which are related to the luminosity distance $D_{L}(z)$ and standard rulers (BAO and CMB) that usually provide information on the large scale of the Universe. In this work, we adopt two different catalogs of data, standard candles and standard rulers, to determine how different samples affect the estimation of cosmological parameters. Here, we turn to a new standard candle compilation of 1598 quasars from X-ray and UV flux measurements with a redshift range $0.036 \leq z \leq 5.1003$ \cite{Risaliti2015}, which has become an effective probe to investigate different cosmological parameters \cite{UV5,Lian2021,xubing2021,khadka2021} especially the cosmic curvature $\Omega_{k}$ \cite{UV2,UV3}, and the cosmic distance duality relation \cite{UV1,liutonghua2020} in the early universe ($z\sim 5$). Besides, the newest SNe Ia sample `` Pantheon" consists of 1048 points
spanning a redshift range $0.01 \leq z \leq 2.3$ \cite{pantheon}, is also adopted in our work as a standard candle. For standard rulers, the angular size from 120 compact radio quasars obtained by very-long baseline interferometry (VLBI) from \cite{caoshuo2017AA,Cao20017qsoas,AS3} is taken into consideration covering the redshift range $0.46 \leq z \leq 2.76$, which has also been widely used in many cosmological analyses, such as the observational constraints on the interaction between cosmic dark sectors \cite{lixiaolei,AS8,AS5}, General Relatively and modified gravity theories \cite{Cao20017qsoas,AS1,xutengpeng2018,AS6}, the Hubble constant and cosmic curvature \cite{AS2,qijingzhao2021}. Additionally, we also adopt 11 BAO data points from BOSS DR12 at $z_{\textrm{eff}}=0.38,0.51,0.61$ \cite{Alam2017}, 6dFGs and SDSS MGS at $z_{\textrm{eff}}=0.122$ \cite{carter2018}, DES Y1 results at $z_{\textrm{eff}}=0.81$ \cite{DES2018}, eBOSS DR14 at $z_{\textrm{eff}}=1.52$ \cite{ata2018} and $z_{\textrm{eff}}=2.34$ \cite{dsa2019}. Specially, introducing quasar measurements to constrain cosmological parameters is beneficial for studying the evolution of cosmological models at higher redshifts \cite{Lian:2021tca,AS7,liutonghua2021}.

In this paper, we focus on standard candles and rulers to constrain four Chaplygin gas cosmological models with the goal of investigating the difference between standard candles and standard rulers and distinguishing these Chaplygin gas models by statistical analysis. This paper is organized as follows. In Section 2, we briefly introduce the basic equations of cosmological models, including GCG, MCG, NGCG, and VGCG. In Section 3, we describe the observational data adopted in this work and perform a Markov chain Monte Carlo (MCMC) analysis using different data sets. The results from observational constraints and the corresponding analysis are displayed in Section 4, as well as some statistical techniques of model comparison presented in Section 5. Finally, our conclusions are summarized in Section 6.
\section{Chaplygin gas cosmologies}
\label{sec:me} 
In this section, we give a description of four types of Chaplygin gas models in a spatially flat universe, including GCG, MCG, NGCG, and VGCG models. Moreover, to obtain stringent constraints on key cosmological parameters, we use the prior on the baryon density parameter $\Omega_{b}$ and radiation density parameter $\Omega_{r}$ from \cite{planck2018result}.

\subsection{GCG model}
The GCG model, which is extended from the CG model, has been generally studied to explain the accelerating universe \cite{Zhangjingfei32,Bento2003,zhu2004,Lixiaolei22,lixiaolei50,lixiaolei,Lian2021}. In this model, the dark energy and dark matter could be unified with an exotic equation, which is introduced as
\begin{equation}
    p_{\mathrm{gcg}}=-\frac{A}{\rho_{\mathrm{gcg}}^{\alpha}},
\label{eq:equation1}
\end{equation}
where $p_{gcg}$ and $\rho_{\mathrm{gcg}}=\rho_{de}+\rho_{dm}$ present the pressure and density of Chaplygin gas, respectively. $A$ is a positive constant and $0 \leq \alpha \leq 1$. When $\alpha=1$, the GCG model reduces to the CG model, and when $\alpha=0$, the GCG model reduces to the $\Lambda$CDM model. The energy density of the GCG model is expressed as
\begin{equation}
    \rho_{\mathrm{gcg}}(a)=\rho_{\mathrm{gcg} 0}\left(A_{\mathrm{s}}+\frac{1-A_{\mathrm{s}}}{a^{3(1+\alpha)}}\right)^{\frac{1}{1+\alpha}},
    \label{eq:equation2}
\end{equation}
where $a$ is a scale factor, which is related to the observable redshift as $a=\frac{1}{1+z}$, $A_{\mathrm{s}} \equiv A / \rho_{\mathrm{gcg}0}^{1+\alpha}$ is a dimensionless parameter, and $\rho_{\mathrm{gcg}0}$ is the present energy value of the GCG density. $A_{s}$ can be written by the effective total matter density $\Omega_{m}$ and $\alpha$ as
\begin{equation}
A_{s}=1-\left(\frac{\Omega_{m}-\Omega_{b}}{1-\Omega_{b}}\right)^{1+\alpha}.
\end{equation}
Therefore, we can derive the normalized Hubble parameter $E(z)$ for this model as
\begin{eqnarray}
E^{2}(z)&=&\Omega_{\mathrm{b}}(1+z)^{3}+\Omega_{\mathrm{r}}(1+z)^{4}+ \nonumber \\
        &&\left(1-\Omega_{\mathrm{b}}-\Omega_{\mathrm{r}}\right)\left(A_{\mathrm{s}}+\left(1-A_{\mathrm{s}}\right)(1+z)^{3(1+\beta)}\right)^{\frac{1}{1+\beta}}.
    \label{eq:equationGCG}
\end{eqnarray}
where $E(z)=H^{2}(z)/H^{2}_{0}$ and the parameter set is $\mathbf{p} \equiv\left(\Omega_{m}, A_{\mathrm{s}}, \alpha, H_{0} \right)$.

\subsection{MCG model}
The MCG model is also a unified dark matter and dark energy model, which is a modification of the GCG model. It has been widely discussed in many perspectives \cite{lixinxu16,lixinxu14,lixinxu15,xulixin2012modified,li2019MCG,debnath2021MCG}. This class of equation of state is expressed as,
\begin{equation}
    \label{eq:equtionMCGeos}
    p_{\mathrm{mcg}}=B \rho_{\mathrm{mcg}}-\frac{A}{\rho_{\mathrm{mcg}}^{\alpha}},
\end{equation}
where $\rho_{\mathrm{gcg}}=\rho_{DE}+\rho_{DM}$, $A$ is a positive constant, $B$ is a free parameter, and $0 \leq \alpha \leq 1$. When $B=0$, this model corresponds to the GCG model, whereas when $A=0$, it reduces to the standard equation of state of a perfect fluid. Especially, it turns to $\Lambda$CDM model with $B=0$ and $\alpha=0$ and it reduces to CG model with $B=0$ and $\alpha=1$. Considering energy conservation, we can obtain the energy density as
\begin{equation}
    \label{eq:equationmcgrho}
    \rho_{\mathrm{mcg}}=\rho_{\mathrm{mcg} 0}\left[A_{s}+\left(1-A_{s}\right) a^{-3(1+B)(1+\alpha)}\right]^{\frac{1}{1+\alpha}},
\end{equation}
where $A_{s}=A/(1+B) \rho_{\mathrm{mcg}0}^{1+\alpha}$, $B\neq-1$ and $\rho_{\mathrm{mcg}0}$ is the present energy value of the MCG density. Therefore, we can rewrite the normalized Hubble parameter $E(z)=H(z)/H_{0}$ for the MCG model as
\begin{eqnarray}
\label{eq:equationmcgEz}
E^{2}(z)&=&\Omega_{b} (1+z)^{3}+\Omega_{r} (1+z)^{4}+(1-\Omega_{b}-\Omega_{r})\times \nonumber \\
&&[A_{s}+(1-A_{s}) (1+z)^{3(1+B)(1+\alpha)}]^{\frac{1}{1+\alpha}}.
\end{eqnarray}
For MCG, the parameter set is $\mathbf{p} \equiv\left(\Omega_{m},A_{\mathrm{s}}, B,\alpha,H_{0} \right)$.

\subsection{NGCG model}
The NGCG model has been studied in previous work, such as \cite{zhangjingfei34,liaokai2013,Zhangjingfei2019,salahedin2020NGCG,Almamon2021}. In the NGCG model, it assumes that the exotic background fluid interpolates between a dust-dominated epoch $\rho \sim a^{-3}$ and a cosmological constant-dominated epoch $\rho \sim a^{-3\left(1+\omega\right)}$, which is portrayed as a unification of X-type dark energy and dark matter. Specifically, when $\omega=-1$, the NGCG model reduces to the GCG model, while $\omega=-1$ and $\alpha=0$, it reduces to the XCDM model. The equation of state of NGCG is given by,
\begin{equation}
    \label{eq:equationngcg_p}
    p_{\mathrm{ngcg}}=-\frac{\tilde{A}(a)}{\rho_{\mathrm{ngcg}}^{\alpha}},
\end{equation}
where $\tilde{A}(a)= -w A a^{-3(1+w)(1+\alpha)}$ is a function of the scale factor, and $\alpha$ is a free parameter spanning 0 to 1. The energy density of the NGCG fluid is
\begin{equation}
    \label{eq:equationngcg_rho}
    \rho_{\mathrm{ngcg}}=\rho_{\mathrm{ngcg}0}a^{-3}\left[1-A_{s}+A_{s} a^{-3 w_{\mathrm{de}}(1+\alpha)}\right]^{\frac{1}{1+a}},
\end{equation}
where $A_{s}=\frac{1-\Omega_{m}}{1-\Omega_{\mathrm{b}}}$. Finally, we can get the form of $E(z)=H(z)/H_{0}$ of the NGCG model,
\begin{eqnarray}
\label{eq:equationNGCG_Ez}
E^{2}(z) &=&\Omega_{\mathrm{b}}(1+z)^{3}+\Omega_{\mathrm{r}}(1+z)^{4}+(1-\Omega_{\mathrm{b}}-\Omega_{\mathrm{r}})(1+z)^{3} \nonumber \\
&& \times [1-\frac{1-\Omega_{m}}{1-\Omega_{\mathrm{b}}-\Omega_{\mathrm{r}}}(1-(1+z)^{3 w(1+\alpha)})]^{\frac{1}{1+\alpha}}.
\end{eqnarray}
Hence, for the NGCG model, the parameter set that we adopt is $\mathbf{p} \equiv\left(\Omega_{m}, \omega, \alpha, H_{0} \right)$.

\subsection{VGCG model}
To tackle the late accelerated expansion of the universe, a hybrid model that consists of a fusion of viscous effects and the features of Chaplygin gas, the VGCG model was studied in \cite{VGCG2006,LiweiVGCG,liwei2015,almada2021VGCG}. This model is able to avoid causality problems that arise when only dissipative fluid is considered and alleviate the blowup in the DM power spectrum for GCG models \cite{almada2021VGCG}. The equation of state of the VGCG model is given by
\begin{equation}
p_{\mathrm{vgcg}}=-A / \rho_{\mathrm{vgcg}}^{\alpha}-\sqrt{3} \zeta \rho_{\mathrm{vgcg}}.
\label{eq:equationvgcg_eos}
\end{equation}
One can obtain the standard $\Lambda$CDM model when $\alpha = 0$ and $\zeta=0$, and this model reduces to the GCG model with $\zeta=0$. Then, we can deduce its energy density as,
\begin{eqnarray}
\rho_{\mathrm{vgcg}} &=&\rho_{\mathrm{vgcg} 0}[\frac{B_{s}}{1-\sqrt{3} \zeta}+(1-\frac{B_{s}}{1-\sqrt{3} \zeta}) \times \nonumber \\
&&  a^{-3(1+\alpha)(1-\sqrt{3} \zeta)}]^{\frac{1}{1+\alpha}},
\label{eq:equationvgcg_rho}
\end{eqnarray}
where $B_{s}=A / \rho_{\mathrm{vgcg}0}^{1+\alpha}$, $0 \leq B_{s} \leq 1$ and $\zeta<\frac{1}{\sqrt{3}}$. The dimensionless Hubble parameter $E(z)=H(z)/H_{0}$ is expressed as
\begin{eqnarray}
\label{eq:equationvgcg_ez}
E^{2}(z) &=& \Omega_{b} (1+z)^{3}+\Omega_{r} (1+z)^{4} + \nonumber \\
&& ( 1 - \Omega_{b} - \Omega_{r}) \times [\frac{B_{s}}{1-\sqrt{3} \zeta} + \nonumber \\ &&(1-\frac{B_{s}}{1-\sqrt{3} \zeta}) (1+z)^{3(1+\alpha)(1-\sqrt{3} \zeta)}]^{\frac{1}{1+\alpha}}.
\end{eqnarray}
It is straightforward that the parameter set of the VGCG model is $\mathbf{p} \equiv ( \Omega_{m},B_{s}, \alpha, \zeta, H_0 )$.

\section{Cosmological observations}
\begin{table}
	\centering
	\begin{tabular}{cccc} 
		\hline
		z & measurement & value & ref\\
		\hline
		0.38 & $D_{M}\left(r_{s, \mathrm{fid}} / r_{s}\right)$ & 1512.39 & \cite{Alam2017}\\
		0.38 & $H(z)(r_{s} / r_{s,fid})$ & 81.2087 & \cite{Alam2017}\\
		0.51 & $D_{M}(r_{s,fid} / r_{s})$ & 1975.22 & \cite{Alam2017}\\
		0.51 & $H(z)(r_{s} / r_{s,fid})$ & 90.9029 & \cite{Alam2017}\\
		0.61 & $D_{M}(r_{s,fid} / r_{s})$ & 2306.08 & \cite{Alam2017}\\
		0.61 & $H(z)(r_{s} / r_{s,fid})$ & 98.9647 & \cite{Alam2017}\\
		0.122 & $D_{V}(r_{s,fid} / r_{s})$ & $539 \pm 17$ & \cite{carter2018}\\
		0.81 & $D_{A}/r_{s}$ & $10.75 \pm 0.43$ & \cite{DES2018}\\
		1.52 & $D_{V}(r_{s,fid} / r_{s})$ & $3843 \pm 147$ & \cite{ata2018}\\
		2.34 & $D_{H}/r_{s}$ & 8.86 & \cite{dsa2019}\\
		2.34 & $D_{M}/r_{s}$ & 37.41 & \cite{dsa2019}\\
		\hline
	\end{tabular}
    \caption{The newest observations of BAO used in this analysis.}
    \label{tab:baodata}
\end{table}

\begin{figure}
    \centering
    \includegraphics[width=\columnwidth]{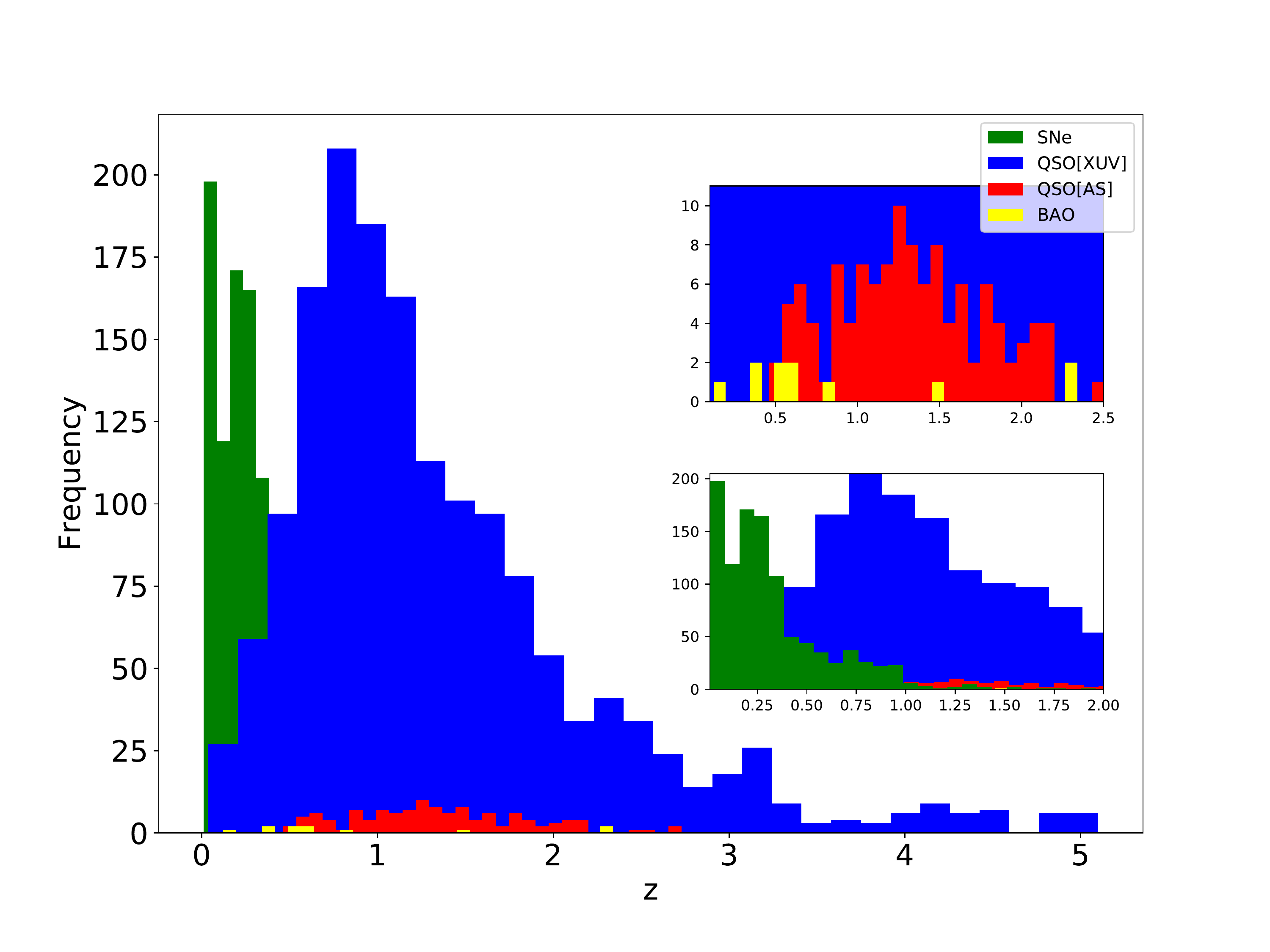}
    \caption{The redshift distribution of the SNe Ia, quasars, and BAO measurements.}
    \label{fig:distribution}
\end{figure}

In this section, we use three catalogs to constrain cosmological models: (1) a standard candle combination of quasars from X-ray and UV flux measurements and SNe Ia samples; (2) a standard ruler set of intermediate-luminosity radio quasars and BAO data listed in Table~\ref{tab:baodata}; and (3) a combination of standard candles and rulers. Additionally, in Fig.~\ref{fig:distribution}, we display the redshift distributions of standard candles and rulers. 

\subsection{QSO[X-ray and UV flux]}

The latest compilation of quasar (QSO[XUV]) from X-ray and UV flux measurements is recognized via the X-ray luminosity and UV luminosity ($L_{X}-L_{UV}$) relation \cite{2019NatAs...3..272R} and used to constrain cosmological model parameters \cite{Risaliti2015,Lian2021}. The $L_{X}-L_{UV}$ relation is given by,
\begin{equation}
\label{eq:LXLUV}
\log (L_{X})=\gamma \log (L_{UV})+\beta,
\end{equation}
where the slopes $\gamma$ and $\beta$ are free parameters that can be measured from the dataset. When we express luminosities in terms of fluxes, $F=L / 4 \pi D_{L}(z)^{2}$, Eq. (\ref{eq:LXLUV}) becomes
\begin{equation}
\label{eq:FXUV}
\log \left(F_{X}\right)=\gamma \log \left(F_{U V}\right)+2(\gamma-1) \log \left(D_{L}\right)+(\gamma-1) \log (4 \pi)+\beta,
\end{equation}
where $F_{X}$ and $F_{UV}$ are the quasar X-ray and UV fluxes, respectively, and $D_{L}$ is the luminosity distance, which is determined via
\begin{equation}
\label{eq:DL}
D_{L}(z, \hat{p})=\frac{c(1+z)}{H_{0}} \int_{0}^{z} \frac{d z^{\prime}}{E\left(z^{\prime}\right)},
\end{equation}
where $E(z)$ depends on different cosmological models.

To obtain the likelihood function, we use Eq. (\ref{eq:FXUV}) and Eq. (\ref{eq:DL}) in a specific model as
\begin{equation}
\mathcal{L}_{F_{X}}=-\frac{1}{2} \sum_{i=1}^{N}\left[\frac{\left[\log \left(F_{X, i}^{\mathrm{obs}}\right)-\log \left(F_{X, i}^{\mathrm{th}}\right)\right]^{2}}{s_{i}^{2}}+\ln \left(2 \pi s_{i}^{2}\right)\right],
\end{equation}
where $\ln =\log _{e}$, $s_{i}^{2}=\sigma_{i}^{2}+\delta^{2}$, and where $\sigma_{i}$ and $\delta$ are the data error on the observed flux and the global intrinsic dispersion, respectively. In addition, according to \cite{2019NatAs...3..272R,khadka2021}, we employ the QSO from X-ray and UV fluxes in the analysis with the chi-square statistic
\begin{equation}
\chi_{F_{x}, m i n}^{2}=-2 \ln (L F)_{\min }-\sum_{i=1}^{1598} \ln \left(2 \pi\left(\sigma_{i}^{2}+\delta_{best-fit}^{2}\right)\right).
\end{equation}

\subsection{SNe Ia}
To use the Pantheon sample, first, we should determine the corresponding observable value and its theoretical value. The observable value given in the Pantheon sample is a corrected magnitude; see Table~A17 of \cite{pantheon} for more details, expressed by
\begin{eqnarray}
Y^{obs} &=& m_B+K \nonumber\\
        &=& \mu+M,
\label{eq:Y_obs}
\end{eqnarray}
where $\mu$ is the distance modulus, $m_B$ is the apparent B-band magnitude, and $M$ is the absolute B-band magnitude of fiducial SNe Ia. There is a correction term $K = \alpha x_1-\beta c+\Delta_M+\Delta_B$ that includes the corrections related to four different sources (for more details, see \cite{pantheon}). The theoretical value is given by,
\begin{eqnarray}
Y^{th}&=& 5\log(D_L)+25 +M \nonumber\\
&=&5\log[(1+z)D(z)]+ Y_0,
\label{eq:Y_th}
\end{eqnarray}
where the constant term $Y_0 = M+5log(\frac{cH_0^{-1}}{Mpc})+25$, which should be marginalized by the methodology presented in \cite{Giostri2019}. The chi-square for the Pantheon sample can be given by
\begin{equation}
\label{eq:chi2SNe}
\chi^{2}_{\textrm{SNe}}={\Delta \overrightarrow{Y}}^T\cdot\textbf{C}^{-1}\cdot{\Delta \overrightarrow{Y}},
\end{equation}
where $\Delta \overrightarrow{Y}_i = [Y^{obs}_i-Y^{th}(z_i; Y_0,\textbf{p})]$ and the covariance matrix $\textbf{C}$ of the sample includes the contributions from both the statistical and systematic errors \cite{pantheon}.

\subsection{QSO[AS]}
\cite{Cao20017qsoas} extracted 120 compact radio quasars (QSO[AS]) based on a 2.29 GHz VLBI all-sky survey of 613 milliarcsecond ultracompact radio sources, covering a redshift range from 0.46 to 2.76. The observable value angular sizes $\theta_{obs}(z)$ is related to the intrinsic length $\ell_m$ and the angular diameter distance $D_A(z)$ \cite{Cao:2015APJ,AS3}. The corresponding theoretical angular size is defined by
\begin{equation}
\theta_{th}(z)=\frac{\ell_{m}}{D_{A}(z)},
\end{equation}
where $\ell_{m}$ is the intrinsic metric linear size, which is calibrated to $11.03 \pm 0.25$ pc by an independent method introduced in \cite{caoshuo2017AA}, and $D_{A}(z)$ is the angular diameter distance 
\begin{equation}
D_{A}(z)=\frac{D_{L}(z)}{(1+z)^{2}},
\end{equation}
where $D_{L}(z)$ is defined by Eq.(\ref{eq:DL}). Therefore, we calculate the chi-square function by
\begin{equation}
\chi_{\textrm{QSO}}^{2}=\sum_{i}^{120} \frac{\left(\theta\left(z_{i} ; \mathbf{p}\right)-\theta_{i}^{o b s}\right)^{2}}{\sigma_{i}^{2}}.
\end{equation}
where $\theta(z_{i};\hat{p})$ is the theoretical value of the angular size and the total uncertainty can be expressed as $\sigma^{2}_{i}=\sigma^{2}_{stat,i}+\sigma^{2}_{sys,i}$.

\subsection{BAO}
The BAO data is also a powerful cosmological probe \cite{eisenstein1998,eisenstein20005}, which is extracted from galaxy redshift surveys. Here, we use 11 BAO measurements summarized in Table~\ref{tab:baodata}. 
The observable quantities used in the measurements are expressed in terms of the transverse co-moving distance $D_M(z)$, the volume-average angular diameter distance $D_V(z)$, the Hubble rate $H(z)\equiv H_{0}E(z)$, the Hubble distance $D_{H}\equiv c/H(z)$, the sound horizon at the drag epoch $r_s$, and its fiducial value $r_{\rm{s,fid}}$. 
In a flat universe, the transverse co-moving distance $D_M(z)$ equals the line-of-sight co-moving distance $D_{C}(z)$, which is expressed as
\begin{equation}
\label{eq:DC}
D_{C}=\frac{c}{H_{0}} \int_{0}^{z} \frac{d z^{\prime}}{E\left(z^{\prime}\right)},
\end{equation}
where $c$ is the velocity of light. The volume-average angular diameter distance is 
\begin{equation}
D_{V}(z)=\left[\frac{c z}{H_{0}} \frac{D_{M}^{2}(z)}{E(z)}\right]^{1 / 3}.
\label{eq:equationDVz}
\end{equation}
Following \cite{ryan2019}, we use the fitting formula of \cite{eisenstein1998} to compute $r_s$ and calculate $r_{s,fid}$ by using the fiducial cosmological model.

Most of data we used are correlated; however, those from \cite{carter2018,DES2018,ata2018}) are uncorrelated. For the uncorrelated data points, the chi-square statistic is expressed as
\begin{equation}
\chi_{\mathrm{BAO}}^{2}(p)=\sum_{i=1}^{N} \frac{\left[A_{\mathrm{th}}\left(p , z_{i}\right)-A_{\mathrm{obs}}\left(z_{i}\right)\right]^{2}}{\sigma_{i}^{2}},
\end{equation}
where $A_{th}(p,z_i)$ denotes the model predictions at the effective redshift, $A_{obs}(z_i)$ is the observational value and $\sigma_i$ is the error bar of the measurements. For the correlated data points from \cite{Alam2017,dsa2019}, it requires
\begin{equation}
\chi_{\mathrm{BAO}}^{2}(p)=\left[\vec{A}_{\mathrm{th}}(p)-\vec{A}_{\mathrm{obs}}\right]^{T} C^{-1}\left[\vec{A}_{\mathrm{th}}(p)-\vec{A}_{\mathrm{obs}}\right],
\label{eq:chi2_BAO}
\end{equation}
where $C^{-1}$ is the inverse of the covariance matrix. The corresponding covariance matrix of \cite{Alam2017} is available from the SDSS website, and that of \cite{dsa2019} is presented in \cite{Caoshulei2020}. 

In the cosmological analysis, the probability distributions of model parameters are obtained with an affine invariant Markov chain Monte Carlo (MCMC) ensemble sampler (emcee) \cite{emcee}, where the statistic can be determined with
\begin{equation}
\mathcal{L}(p)=e^{-\frac{\chi(p)^{2}}{2}},
\end{equation}
where $p$ is the set of model parameters from different cosmological models. 

\section{Results and discussion}

\begin{table*}
    \caption{The best-fit values and 68\% confidence limits for the CG cosmological parameters in each model (GCG, MCG, NGCG, and VGCG) and data set (QSO[XUV]+SNe Ia, QSO[AS]+BAO, and QSO[XUV]+SNe Ia+QSO[AS]+BAO).}
\resizebox{\textwidth}{!}{
\begin{tabular}{ccccccc}
\hline\noalign{\smallskip}
    Model & Data & $\Omega_{m}$ & $A_{s}$ & $\alpha$ & $H_{0}$ (km/s/Mpc)   \\
\noalign{\smallskip}\hline\noalign{\smallskip}
    GCG & QSO[XUV]+SNe Ia & $0.53^{+0.53}_{-0.29}$ & $0.78 \pm 0.06$ & $0.46^{+0.57}_{-0.42}$ & $68.27^{+6.98}_{-4.76}$\\
        & QSO[AS]+BAO & $0.33\pm0.02$ & $0.60 \pm 0.10$ & $-0.33^{+0.27}_{-0.24}$ & $65.81^{+2.26}_{-2.28}$ \\
        & Combination & $0.31\pm 0.01$ & $0.73 \pm 0.04$ & $0.03^{+0.17}_{-0.14}$ & $68.26^{+1.18}_{-1.08}$\\
\noalign{\smallskip}\hline\noalign{\smallskip}

    Model & Data & $\Omega_{m}$ & $A_{s}$ & $B$ & $\alpha$ & $H_{0}$ (km/s/Mpc)\\
\noalign{\smallskip}\hline\noalign{\smallskip}

    MCG & QSO[XUV]+SNe Ia & $0.47^{+0.36}_{-0.27}$ & $0.81^{+0.06}_{-0.09}$ & $0.12^{+0.26}_{-0.21}$ & $0.20^{+0.58}_{-0.39}$ & $68.28^{+7.23}_{-3.48}$ \\
        & QSO[AS]+BAO &$0.33 \pm 0.02$ & $0.61^{+0.10}_{-0.14}$ & $-0.12^{+0.18}_{-0.09}$ & $0.05^{+0.87}_{-0.52}$ & $66.27^{+2.01}_{-2.30}$ \\
        & Combination & $0.31\pm 0.01$ & $0.73^{+0.04}_{-0.06}$ & $-0.14^{+0.13}_{-0.06}$ & $0.71^{+0.78}_{-0.71}$ & $68.09^{+1.11}_{-1.06}$ \\
\noalign{\smallskip}\hline\noalign{\smallskip}

    Model & Data & $\Omega_{m}$ & $\omega$ & $\alpha$ & $H_{0}$ (km/s/Mpc)\\
\noalign{\smallskip}\hline\noalign{\smallskip}

    NGCG & QSO[XUV]+SNe Ia & $0.30^{+0.16}_{-0.14}$ & $-1.10^{+0.24}_{-0.39}$ & $0.23^{+0.89}_{-0.53}$ & $68.50^{+7.19}_{-5.21}$\\
         & QSO[AS]+BAO & $0.34 \pm 0.02$ & $-0.79^{+0.13}_{-0.14}$ & $-0.12^{+0.12}_{-0.17}$ & $65.16^{+2.38}_{-2.18}$\\
         & Combination & $0.31 \pm 0.01$ & $-1.01^{+0.05}_{-0.06}$ & $0.01^{+0.09}_{-0.08}$ & $68.34^{+1.19}_{-1.09}$\\
\noalign{\smallskip}\hline\noalign{\smallskip}

    Model & Data & $\Omega_{m}$ & $B_{s}$ & $\alpha$ & $\zeta$ & $H_{0}$ (km/s/Mpc) \\
 \noalign{\smallskip}\hline\noalign{\smallskip}

    VGCG & QSO[XUV]+SNe Ia & $0.47^{+0.36}_{-0.31}$ & $0.82^{+0.13}_{-0.19}$ & $0.41^{+0.74}_{-0.49}$ & $-0.02^{+0.10}_{-0.08}$ & $68.58^{+6.82}_{-5.16}$\\
         & QSO[AS]+BAO & $0.33 \pm 0.02 $ & $0.55^{+0.17}_{-0.16}$ & $0.08^{+0.99}_{-0.57}$ & $0.07^{+0.06}_{-0.12}$ & $66.32^{+2.16}_{-2.42}$\\
         & Combination & $0.31 \pm 0.01$ & $0.64^{+0.10}_{-0.07}$ & $0.61^{+0.82}_{-0.66}$ & $0.07^{+0.04}_{-0.08}$ & $68.21^{+1.19}_{-1.04}$\\ 
\noalign{\smallskip}\hline
    \end{tabular}}

    \label{tab:results}
\end{table*}

\begin{table}
    \caption{The values of DIC and their differences for CG and $\Lambda$CDM cosmologies. The Jensen-Shannon divergence between $\Lambda$CDM and other cosmological models is also calculated with respect to $\Omega_{m}$ and $H_{0}$.}
    \label{tab:IC}
\resizebox{\columnwidth}{!}{
\begin{tabular}{cccccc}
\hline\noalign{\smallskip}
     Data & Model & DIC & $\Delta$DIC & $D_{JS}(\Omega_{m})$ & $D_{JS}(H_{0})$\\
\noalign{\smallskip}\hline\noalign{\smallskip}
    QSO[XUV]+SNe Ia & $\Lambda$CDM & 2632.34 & 0 & 0 & 0\\
    &GCG & 2635.10 & 2.76 & 0.721 & 0.124\\
    & MCG & 2625.76 & -6.58 & 0.722 & 0.510\\
    & NGCG & 2636.31 & 3.97 & 0.681 & 0.083 \\
    & VGCG & 2638.69 & 6.36 & 0.727 & 0.092\\
\noalign{\smallskip}\hline\noalign{\smallskip}
    QSO[AS]+BAO & $\Lambda$CDM & 616.42 & 0 & 0 & 0 \\
    & GCG & 618.29 & 1.87 & 0.296 & 0.710 \\
    & MCG & 608.04 & -8.38 & 0.300 & 0.662 \\
    & NGCG & 617.97 & 1.55 & 0.468 & 0.927 \\
    & VGCG & 605.54 & -10.88 & 0.319 & 0.655 \\
\noalign{\smallskip}\hline\noalign{\smallskip}
    Combination & $\Lambda$CDM  & 3245.86 & 0 & 0 & 0\\ 
    & GCG  & 3252.10 & 6.24 &0.074& 0.286\\
    & MCG  & 3246.42 & 0.56 & 0.132 & 0.275\\
    & NGCG & 3251.15 & 5.29 &0.077 & 0.288\\
    & VGCG & 3239.22 & -6.64 & 0.131 & 0.277\\
\noalign{\smallskip}\hline
    \end{tabular}}
\end{table}

\begin{figure}
	\includegraphics[width=\columnwidth]{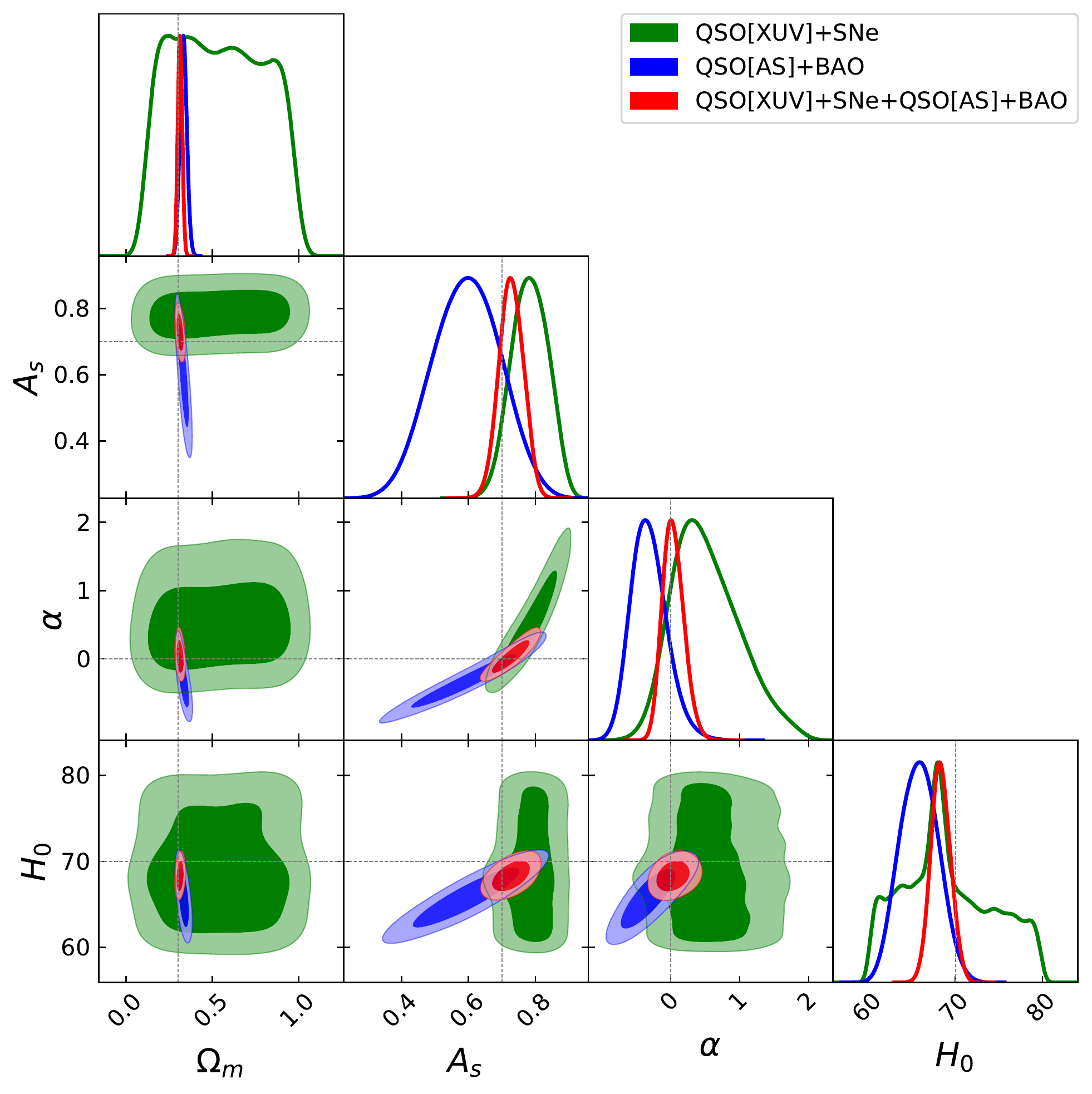}
    \caption{The 1D and 2D probability distributions of model parameters in the GCG model, based on the QSO[XUV]+SNe Ia(green), QSO[AS]+BAO (blue), and a joint sample (red). The contours correspond to 68\% and 95\% confidence levels. The grey line indicates the values of the model parameters that can be recovered to the $\Lambda$CDM scenario, with the fiducial value of $\Omega_{m}=0.30$, $\omega=-1$, $\alpha=0$ and $H_{0}$=70 km/s/Mpc.}
    \label{fig:GCG}
\end{figure}

\begin{figure}
    \centering
    \includegraphics[width=\columnwidth]{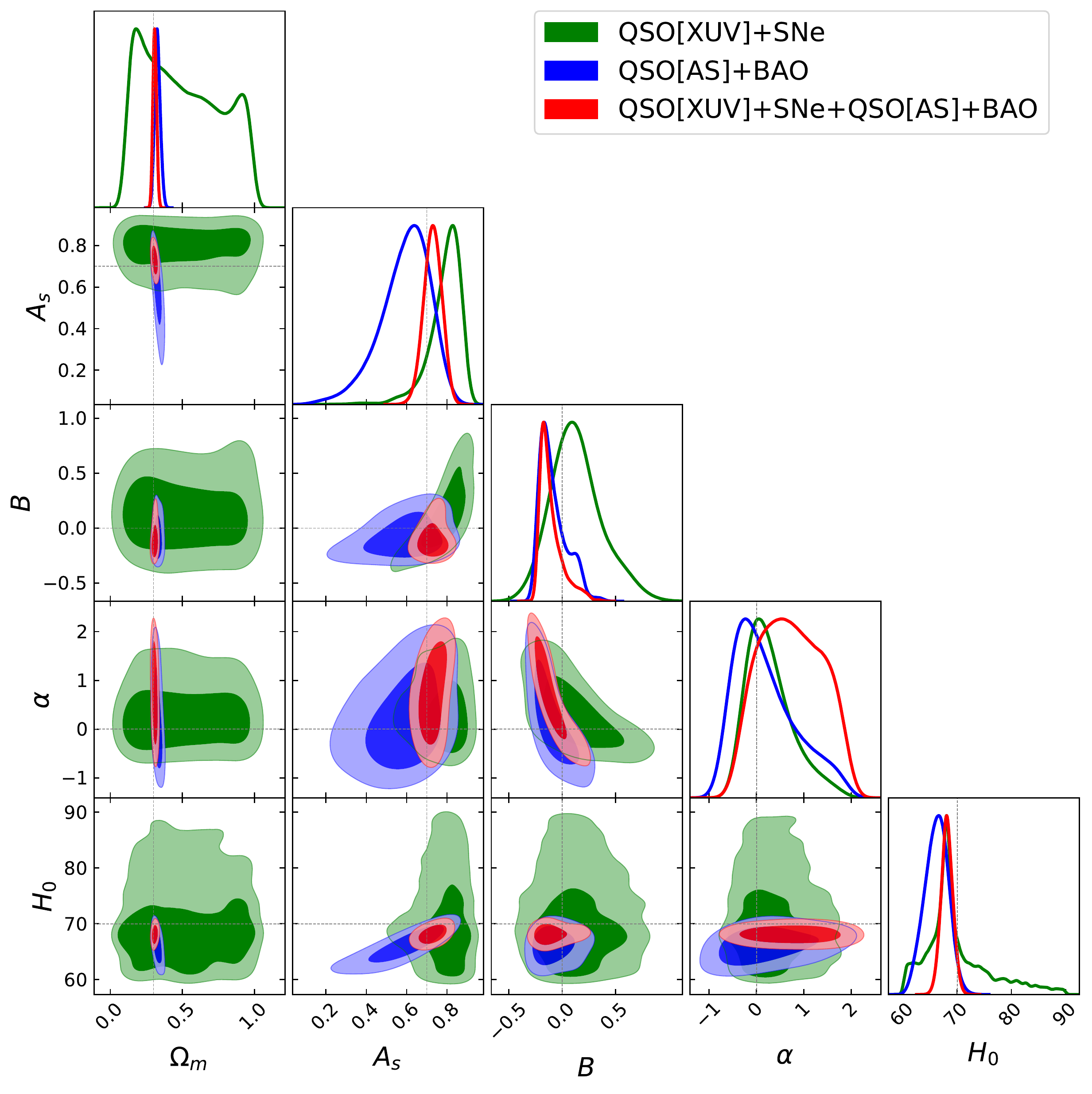}
    \caption{The 1D and 2D probability distributions of model parameters in the MCG model.}
    \label{fig:MCG}
\end{figure}

\begin{figure}
    \centering
    \includegraphics[width=\columnwidth]{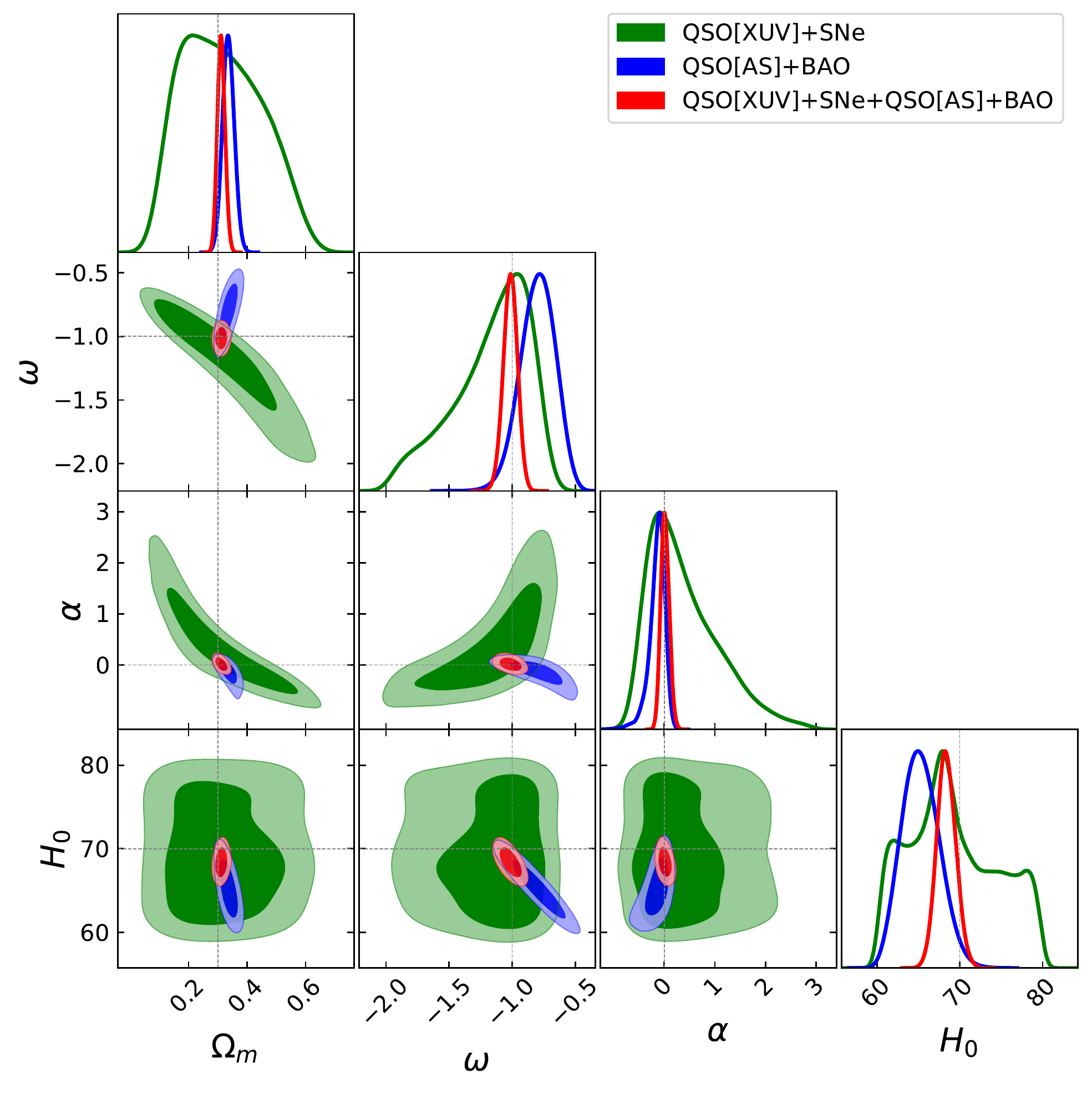}
    \caption{The 1D and 2D probability distributions of model parameters in the NGCG model.}
    \label{fig:NGCG}
\end{figure}

\begin{figure}
    \centering
    \includegraphics[width=\columnwidth]{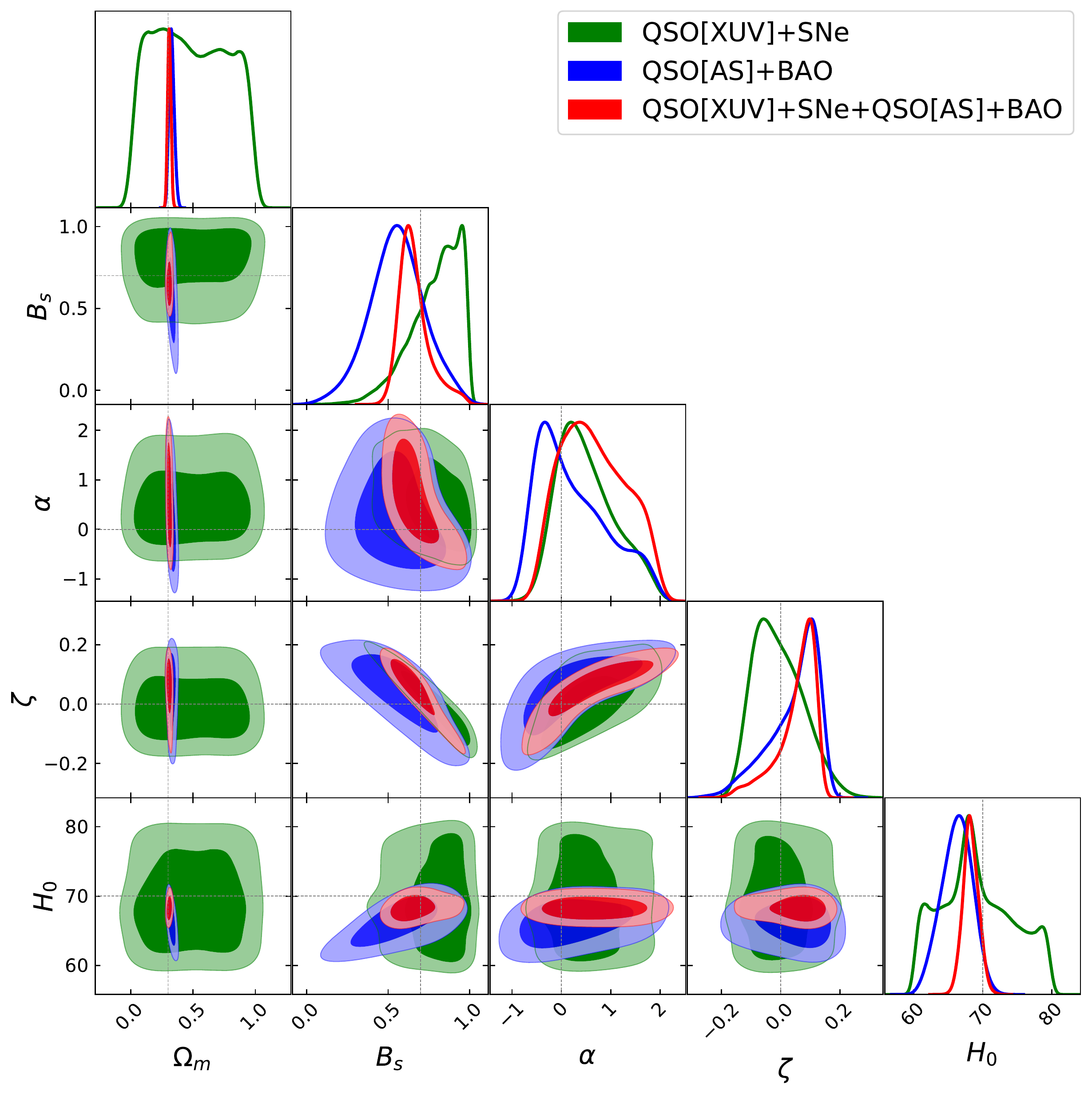}
    \caption{The 1D and 2D probability distributions of model parameters in the VGCG model.}
    \label{fig:VGCG}
\end{figure}

In this section, we display and discuss the constraint results of the cosmological parameters by using the standard candles and rulers data. It shows that how the different types of observational data could inflect the constraints of cosmological parameter estimation. 

\subsection{GCG model}
We present the 1D probability distributions and 2D contours with 1$\sigma$ and 2$\sigma$ confidence levels (CLs) for the GCG model in Fig.~\ref{fig:GCG} and list the best-fit parameters at the 1$\sigma$ confidence level in Table~\ref{tab:results}. 
The standard candle data gives $\Omega_{m}=0.53^{+0.53}_{-0.29}$, $A_{s}=0.78\pm 0.06$,$\alpha=0.46^{+0.57}_{-0.42}$ and $H_{0}=68.27^{+6.98}_{-4.76}$ km/s/Mpc, while the standard ruler data obtains $\Omega_{m}=0.33\pm0.02$, $A_{s}=0.60 \pm 0.10$, $\alpha= -0.33^{+0.27}_{-0.24}$ and $H_{0}=65.81^{+2.26}_{-2.28}$ km/s/Mpc. First, it is clear that the value of $\Omega_{m}$ obtained from standard candles shows a deviation from the Planck collaboration ($\Omega_{m}=0.3103 \pm0.0057$) \cite{planck2018result}. This is because a larger value of the matter density parameter is favored by the recent QSO[XUV] compilation in most cosmological models at higher redshifts ($2.5 < z < 5$), which has been discussed in previous work \cite{Risaliti2015,khadka2020_1}. In addition, $\alpha$ is an important parameter $0 \leq \alpha \leq 1$, where $\alpha=0$ denotes the $\Lambda$CDM model and $\alpha=1$ denotes the CG model. Although previous studies showed that the CG model is ruled out by observations, we find that the CG model is accepted by QSO[XUV]+SNe Ia at a 68\% CL. In addition, the standard ruler data favor the $\Lambda$CDM model at a 95\% CL, as well as the combination sample. For the Hubble constant, our constraint results are in good agreement with the Planck collaboration ($H_{0}=67.66 \pm 0.42$ km/s/Mpc) \cite{planck2018result}, although the values obtained from standard rulers are lower than that from other probes. In addition, the standard ruler data could bring down the error bars of $\Omega_{m}$, $\alpha$ and $H_{0}$ compared with the standard candles. This indicates that QSO[AS]+BAO could give a more restrictive constraint on cosmological parameters. Moreover, it is necessary to refer to the previous results, such as $A_{s}=0.70^{+0.16}_{-0.17}$ and $\alpha=-0.09^{+0.54}_{-0.33}$ constrained from the X-ray gas mass fraction, Type Ia supernovae and Type IIb radio galaxies in \cite{zhu2004} and $\alpha=-0.14^{+0.30}_{-0.19}$ obtained by SNe Ia+H(z)+CMB in \cite{wupuxun2007}, which are consistent with our results from combination data and favor the standard $\Lambda$CDM model. It is worth mentioning that \cite{Lian2021} gave $\Omega_{m}=0.416^{+0.088}_{-0.068}$, $\alpha=2.360^{+1.803}_{-1.793}$ and $H_{0}=69.254^{+4.427}_{-4.970}$ km/s/Mpc from the QSO[XUV]+QSO[AS], which is in good agreement with our results from QSO[XUV]+SNe Ia and includes the CG model at a 68\% CL. This suggests that the latest QSO compilation from X-ray and UV flux measurements slightly favors the CG model and prefers a larger value of $\Omega_{m}$.

\subsection{MCG model}
In the case of the MCG model, the results are presented in Fig.~\ref{fig:MCG} and Table~\ref{tab:results}. The standard candle data generates $\Omega_{m}=0.47^{+0.36}_{-0.27}$, $A_{s}=0.81^{+0.06}_{-0.09}$, $B=0.12^{+0.26}_{-0.21}$, $\alpha=0.20^{+0.58}_{-0.39}$ and $H_{0}=68.28^{+7.23}_{-3.48}$ km/s/Mpc, while the standard ruler data provides $\Omega_{m}=0.33 \pm 0.02$, $A_{s}=0.61^{+0.10}_{-0.14}$, $B=-0.12^{+0.18}_{-0.09}$, $\alpha=0.05^{+0.87}_{-0.52}$ and $H_{0}=66.27^{+2.01}_{-2.30}$ km/s/Mpc. The value of $\Omega_{m}$ obtained from QSO[XUV]+SNe Ia is still higher than that from other probes, which is the same as the case of the GCG model and still consistent with that from \cite{planck2018result} at a 68.3\% CL. In the framework of the MCG model, considering the fact that the parameter $B$ reflects the deviation from the GCG model (the MCG model reduces to the GCG model when $B=0$), the GCG model is accepted by current observations at a 95\% CL in all cases. However, the MCG model shows a tiny deviation from the GCG model by the combination sample at a 68\% CL. For the key parameter $\alpha$ that quantifies the deviation from the CG model and $\Lambda$CDM model, it is clear that the $\Lambda$CDM model, $B=0$ and $\alpha=0$, is accepted by standard candles and standard rulers at 68\% CLs, while the CG model, $B=0$ and $\alpha=1$, is favored by the combination sample at a 95\% CL. However, in the case of the combination sample, both the $\Lambda$CDM and CG models are favored within a 68\% CL.
In other words, this suggests that the $\Lambda$CDM model is more favored by standard candles and standard rulers, respectively, but the CG model is slightly preferred by the combination. Focusing on the Hubble constant, the constraint results agree well with Planck collaboration \cite{planck2018result}. Moreover, our results from combination samples are consistent with the results obtained from SNe Ia+BAO+CMB \cite{xulixin2012modified} with $\alpha=0.000727^{+0.00142}_{-0.00140}$, $B_{s}=0.782^{+0.0163}_{-0.0162}$ and $B=0.000777^{+0.000201}_{-0.000302}$ and from H(z)+BAO+CMB+SNe Ia \cite{Thakur2019} with $\Omega_{m}=0.284^{+0.013}_{-0.014}$, $\alpha=0.046^{+0.107}_{-0.102}$ and $B=0.0026\pm0.005$ at the 1$\sigma$ confidence level. This proves that most cosmological probes favor the $\Lambda$CDM model; however, the inclusion of the QSO sample from X-ray and UV flux measurements \cite{Risaliti2015} at higher redshifts changes to slightly favor the CG model.

\subsection{NGCG model}
In Fig.~\ref{fig:NGCG} and Table~\ref{tab:results}, we show the constraint results of the NGCG model. Compared with the standard candle dataset with $\Omega_{m}=0.30^{+0.16}_{-0.14}$, $\omega=-1.10^{+0.24}_{-0.39}$, $\alpha=0.23^{+0.89}_{-0.53}$ and $H_{0}=68.50^{+7.19}_{-5.21}$ km/s/Mpc, the standard ruler data obtains $\Omega_{m}=0.34\pm0.02$, $\omega=-0.79^{+0.13}_{-0.14}$, $\alpha=-0.12^{+0.12}_{-0.17}$ and $H_{0}=65.16^{+2.38}_{-2.18}$ km/s/Mpc. The most notable thing is that $\Omega_{m}=0.30^{+0.16}_{-0.14}$ from standard candles is consistent with Planck collaboration ($\Omega=0.3103\pm0.0057$) \cite{planck2018result}, however this is contrary in the scenarios of the GCG, MCG and VGCG models. \cite{khadka2020_1} constrained $\Omega_{m} \sim 0.3$ in the XCDM model from only compiled X-ray and UV flux measurements of 1598 quasars, while $\Omega_{m} \sim 0.5-0.6$ in the $\Lambda$CDM and $\phi$CDM models. There are similarities between the NGCG and XCDM models because the parameter $\omega$ in the NGCG model is proposed by a similar idea to that in the XCDM model. Hence, we obtain a normal value of $\Omega_{m}$ in the framework of NGCG, which indicates that X-ray and UV flux measurements of 1598 quasar compilations could help to determine the dark energy and dark matter. 
It should be noted that $\omega$ is a free constant and \cite{NGCG2006} proposed the probability that dark energy behaves in a quintessence-like form with $\omega > -1$ and phantom-like form with $\omega <$ $-1$. The 1$\sigma$ range $\omega \in (-1.05,-0.95)$ from the combination sample implies that there is an equal chance that dark energy behaves as a quintessence-like form or phantom-like form. In all cases, it suggests that the GCG model (i.e., $\omega=-1$) and XCDM model (i.e., $\omega=-1$ and $\alpha=0$) are still supported by the observational data at a 95\% CL. In addition, it is remarkable that the CG model, $\omega=-1$ and $\alpha=1$, is accepted by the standard candle data at a 68\% CL. The Hubble constant obtained in our analysis is more consistent with the results of \cite{planck2018result} at a 68\% CL. Furthermore, we make a comparison with the previous findings in the literature. For instance, \cite{liaokai2013} derived $\Omega_{de}=0.7297^{+0.0229}_{-0.0276}$, $\omega=-1.0510^{+0.1563}_{-0.1685}$ and $\eta=1+\alpha=1.0117^{+0.0469}_{-0.0502}$ with SNe Ia+BAO+WMAP+H(z) data; \cite{Zhangjingfei2019} obtained $\Omega_{de}=0.6879 \pm 0.0078$, $\omega=-1.02 \pm 0.045$, $\alpha=-0.0029 \pm 0.0097$ and $H_{0}=67.78 \pm 0.87$ km/s/Mpc with a joint sample of SNe Ia+BAO+CMB; and \cite{salahedin2020NGCG} stated $\Omega_{m}=0.2508^{+0.0081}_{-0.0097}$, $\omega=-1.041\pm0.045$, $A_{s}=0.7371^{+0.0097}_{-0.0086}$, $\eta=1+\alpha=0.9443\pm0.0097$ and $H_{0}=70.15\pm0.84$ km/s/Mpc with SNe Ia+BAO+CMB+BBN+H(z) data. This indicates that the value of $\Omega_{m}$ from current observations, i.e., SNe Ia, BAO, CMB and H(z), is generally smaller than $\Omega_{m}=0.3103 \pm 0.057$ from \cite{planck2018result}; however, the inclusion of QSO[XUV] and QSO[AS] changes the value of $\Omega_{m}$ to 0.3-0.34 in our work. It indicates that the inclusion of quasar data could help us to study dark matter and dark energy.

\subsection{VGCG model}
The best-fit values for the VGCG model from different observations are shown in Fig.~\ref{fig:VGCG} and Table~\ref{tab:results}. The standard candle data obtains $\Omega_{m}=0.47^{+0.36}_{-0.31}$, $B_{s}=0.82^{+0.13}_{-0.19}$, $\alpha=0.41^{+0.74}_{-0.49}$, $\zeta=-0.0017^{+0.10}_{-0.08}$ and $H_{0}=68.58^{+6.82}_{-5.16}$ km/s/Mpc, while the standard ruler data shows $\Omega_{m}=0.33 \pm 0.02$, $B_{s}=0.55^{+0.17}_{-0.16}$, $\alpha=-0.08^{+0.99}_{-0.57}$, $\zeta=0.07^{+0.06}_{-0.12}$ and $H_{0}=66.32^{+2.16}_{-2.42}$ km/s/Mpc. It indicates that the value of $\Omega_{m}$ is still larger than that of \cite{planck2018result}, since the QSO[XUV] data favors higher $\Omega_{m}$ in most dark energy models \cite{Risaliti2015,khadka2020_1}.
$\zeta$ is the viscosity term that affects the CMB power spectrum about the matter density on the height of the acoustic peaks. From the results shown in Table~\ref{tab:results}, it implies that $\zeta$ is very small, which could alleviate the oscillations causing the blowup in the DM power spectrum in the GCG models. Moreover, the GCG model (i.e., $\zeta=0$) is still favored by the available observations.
On the other hand, $\alpha$ is an important parameter that reflects the deviation from the CG model and $\Lambda$CDM model. In all cases, the CG model cannot be ruled out by current observations at a 68\% CL, while $\Lambda$CDM is still accepted by the observations at a 68\% CL. In other words, it indicates that QSO[XUV], SNe Ia, QSO[AS] and BAO data could not give accurate constraints on $\alpha$. 
In addition, our results on the Hubble constant approve the value of Planck collaboration ($H_{0}=67.66\pm 0.42$ km/s/Mpc) \cite{planck2018result} at a 68\% CL.
It is reasonable to compare to previous studies, such as \cite{liwei2013VGCG} declared $\zeta=0.000708^{+0.00151}_{-0.00155}$ from SNe Ia+BAO+WMAP and \cite{LiweiVGCG} announced $\zeta=0.0000138^{+0.00000614}_{-0.0000105}$ from SNLS3
+BAO+HST. In a recent work, \cite{almada2021VGCG} used a joint sample of SLS+SNe Ia
+BAO+OHD+HIIG and obtained $B_{s}=0.50^{+0.05}_{-0.06}$, $\alpha=0.99^{+0.61}_{-0.58}$, $\zeta=0.13^{+0.02}_{-0.03}$ and $h=0.69\pm0.01$. They concluded that the GCG model, $\zeta=0$, is disfavored by SLS+SNe Ia+BAO+OHD+HIIG at a 68\% CL, which is different from our results with $\zeta=0.07^{+0.04}_{-0.08}$. Moreover, we find that the inclusion of cosmic microwave background data could give a more precise constraint on $\zeta$. 

From our constraint results on the matter density parameter $\Omega_{m}$ in different CG models, it is clear that the standard candle data combining QSO[XUV] with SNe Ia prefers larger values of $\Omega_{m}$ ranging from $0.47-0.53$ except the NGCG model. In \cite{khadka2020_1}, it states that the QSO[XUV] data at $z\sim2-5$ prefers larger values of $\Omega_{m} \sim 0.5-0.6$.
Other studies have concentrated on exploring the tension between high redshift quasar measurements and other observations, such as BAO measurements in \cite{Risaliti2015,razaei2020,yangtao2020cosmography,lixiaolei2021hubble}. It implies that there is an unknown systematic error in the high redshift observations or a stimulus of the new physics and astronomy. Therefore, more accurate cosmological probes are required to solve the problem of the $\Omega_{m}$ inconsistency from high and low redshift observations. On the other hand, it is also rewarding to comment on the possible alleviation of the $H_0$ tension by the VGCG and MCG model. Based on our results presented in Table~\ref{tab:results}, the constraint on the Hubble constant lies in the range of $H_{0}=68.28^{+7.23}_{-3.48}$ km/s/Mpc to $H_{0}=68.58^{+6.82}_{-5.16}$ km/s/Mpc for the standard candles, as well as $H_{0}=68.09^{+1.11}_{-1.06}$ km/s/Mpc to $H_{0}=68.21^{+1.19}_{-1.04}$ km/s/Mpc for the combined sample. It is noteworthy that these two Chaplygin gas models suggest a central value of the Hubble constant between the Planck experiment \cite{planck2018result}, $H_{0}=67.4 \pm 0.5$ km/s/Mpc and the SH0ES experiment \cite{riessH0}, $H_{0}=74.03 \pm 1.42$ km/s/Mpc.

\section{Statistical analysis}

The statistical analysis is essential to diagnose the different models. Hence, we apply the Jensen-Shannon Divergence, statefinder diagnostic and the deviance information criterion. In this section, we compare these models and discuss how strongly are they favored by the observational data sets. 

\label{sec:stat}
\subsection{Jensen-Shannon Divergence}
\label{sec:jsd}

\begin{figure}
\centering
\subfigure{\includegraphics[width=2.5in]{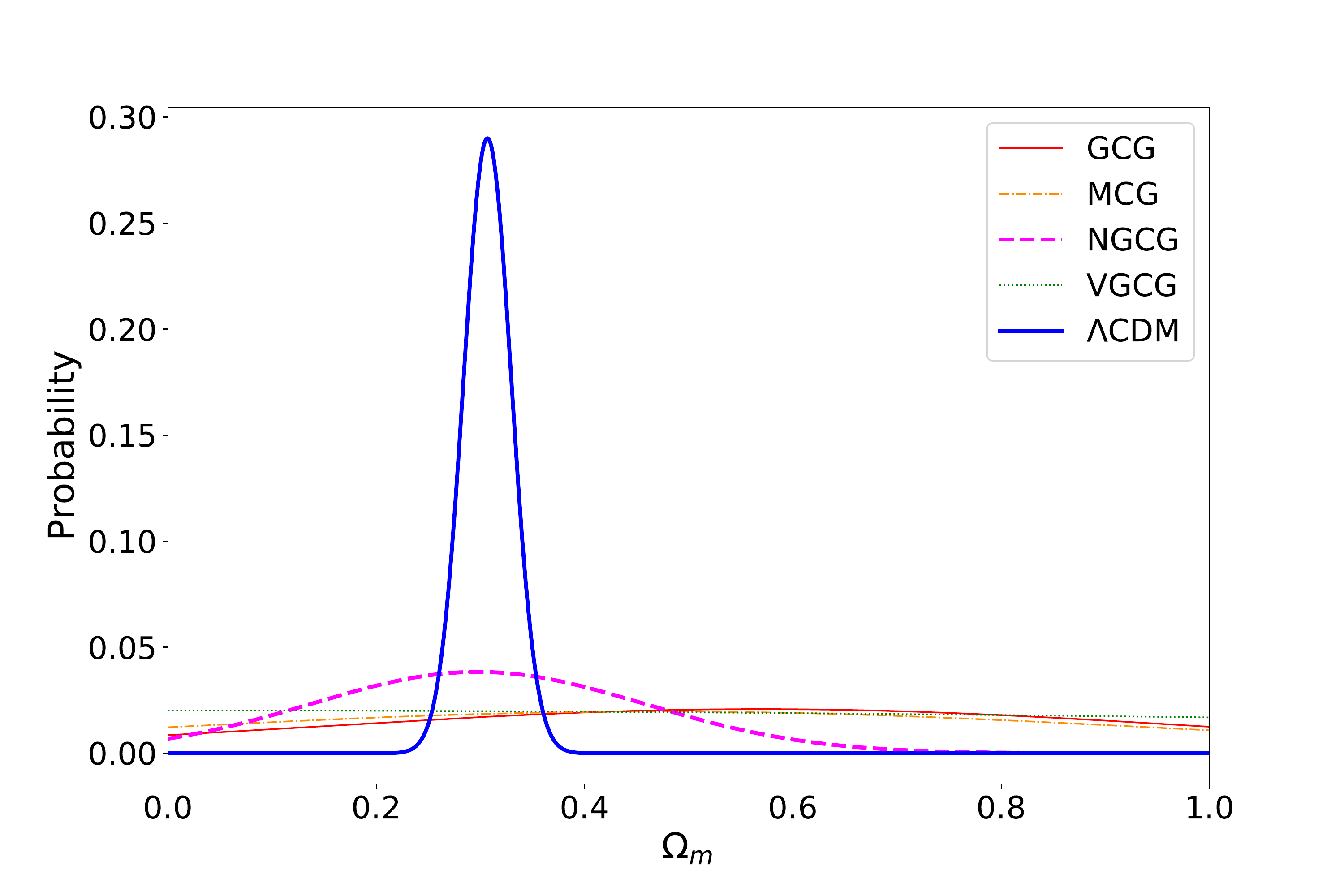}}
\subfigure{\includegraphics[width=2.5in]{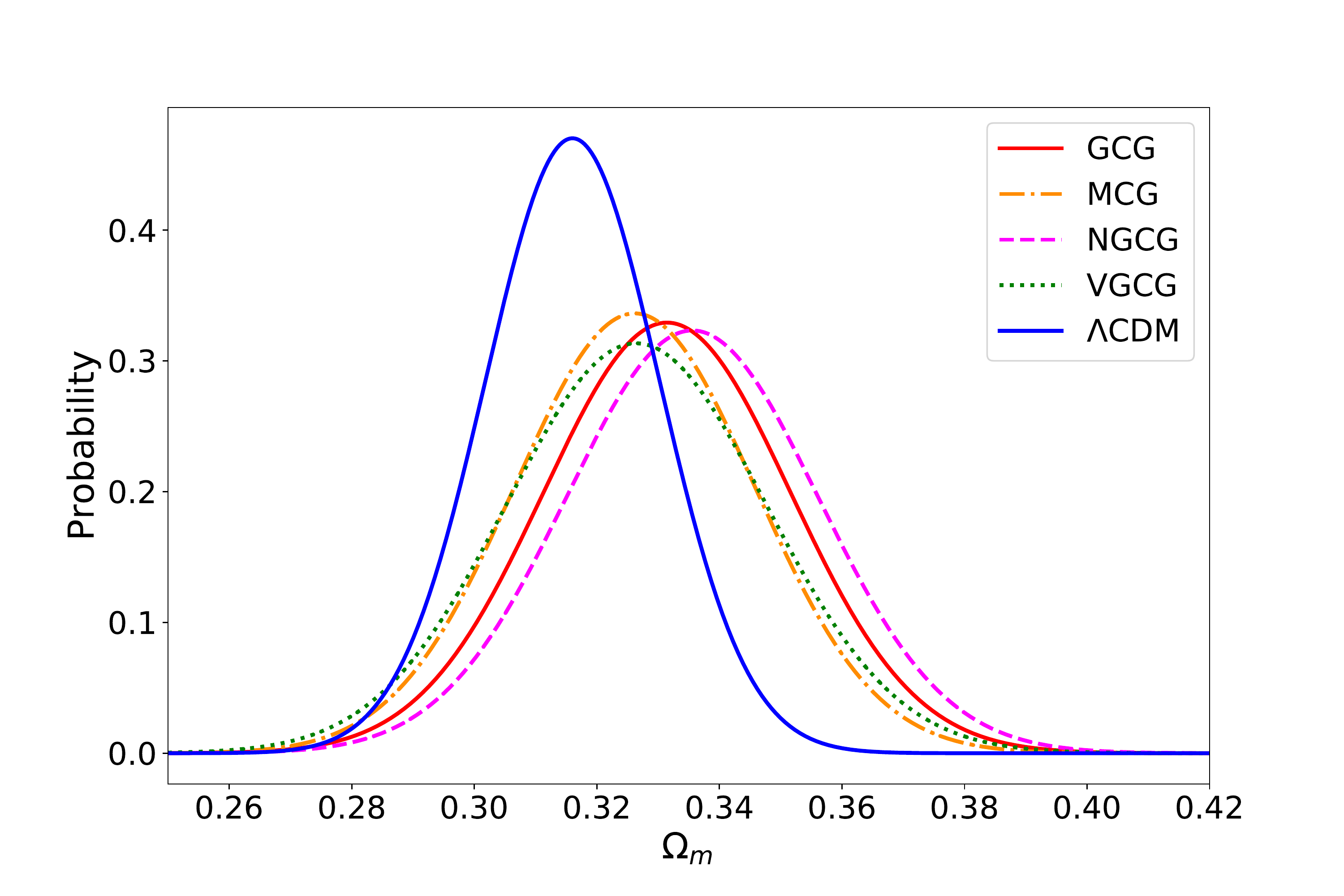}}
\subfigure{\includegraphics[width=2.5in]{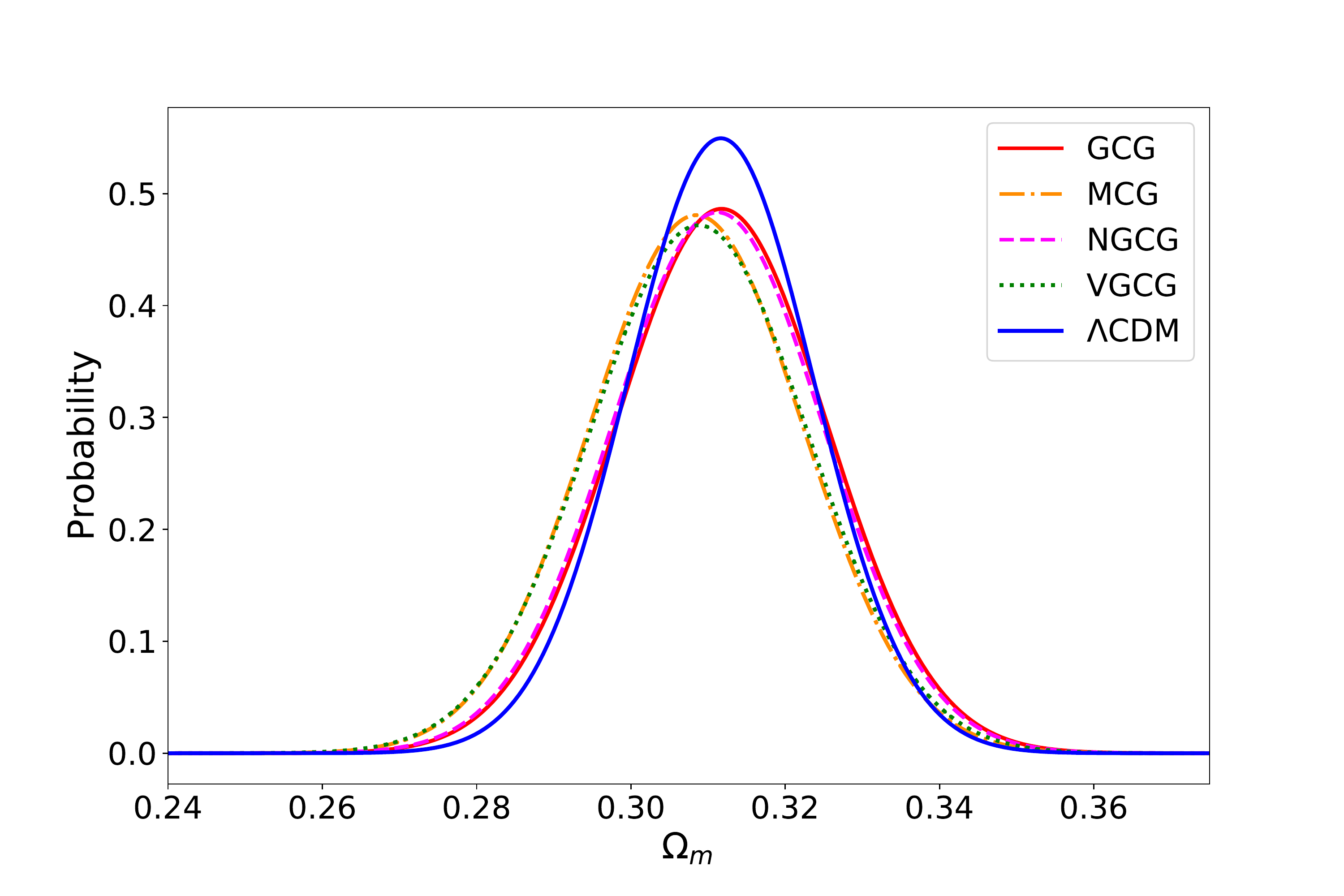}}
\caption{The posterior distributions of $\Omega_{m}$ for the GCG, MCG, NGCG, VGCG and $\Lambda$CDM models, with the standard candles, standard rulers and combination data from the top to the bottom. We adopt the posterior distributions of $\Omega_{m}$ from Table~\ref{tab:results}.}
\label{fig:JSD_Om}
\end{figure}

\begin{figure}
\centering
\subfigure{\includegraphics[width=2.5in]{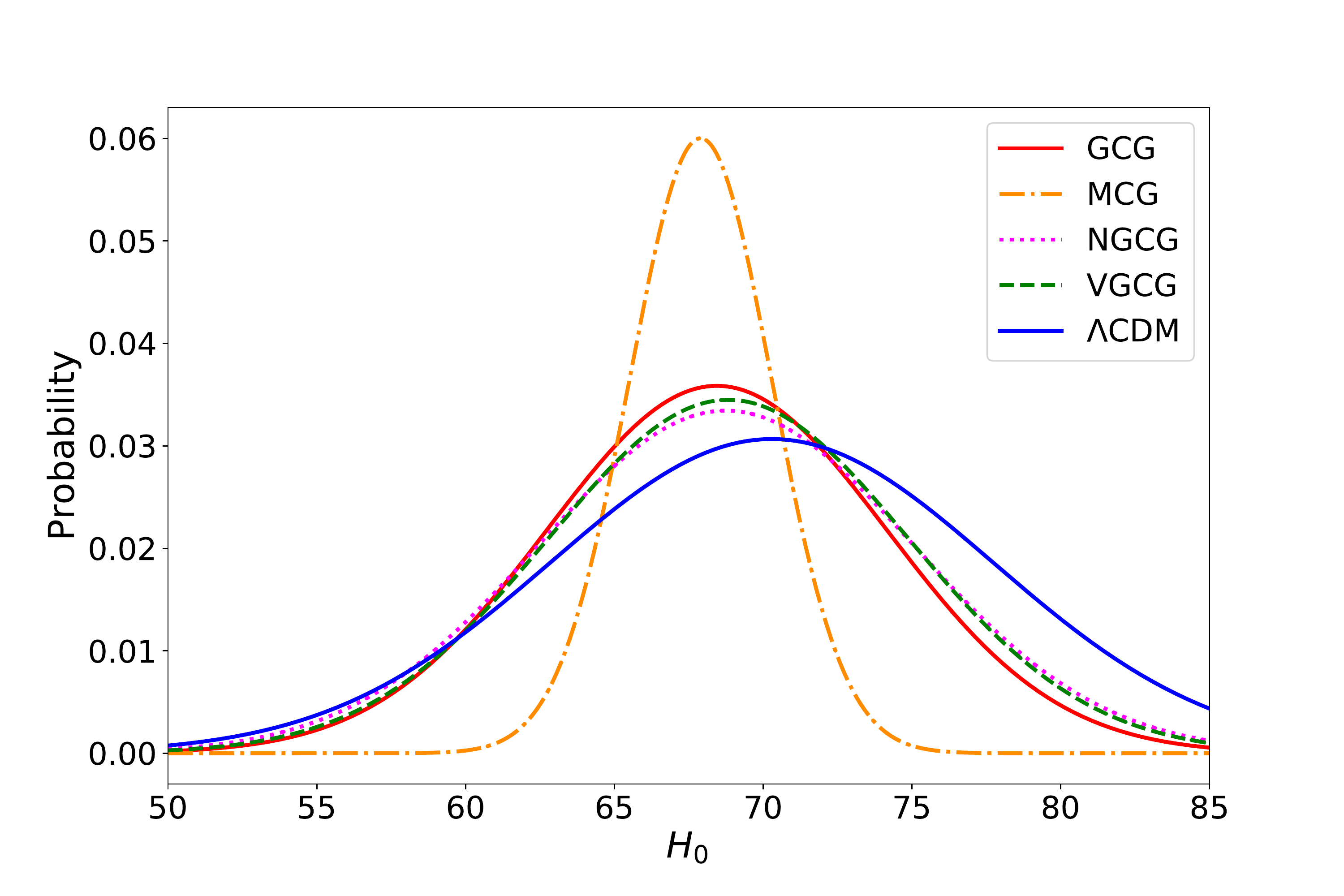}}
\subfigure{\includegraphics[width=2.5in]{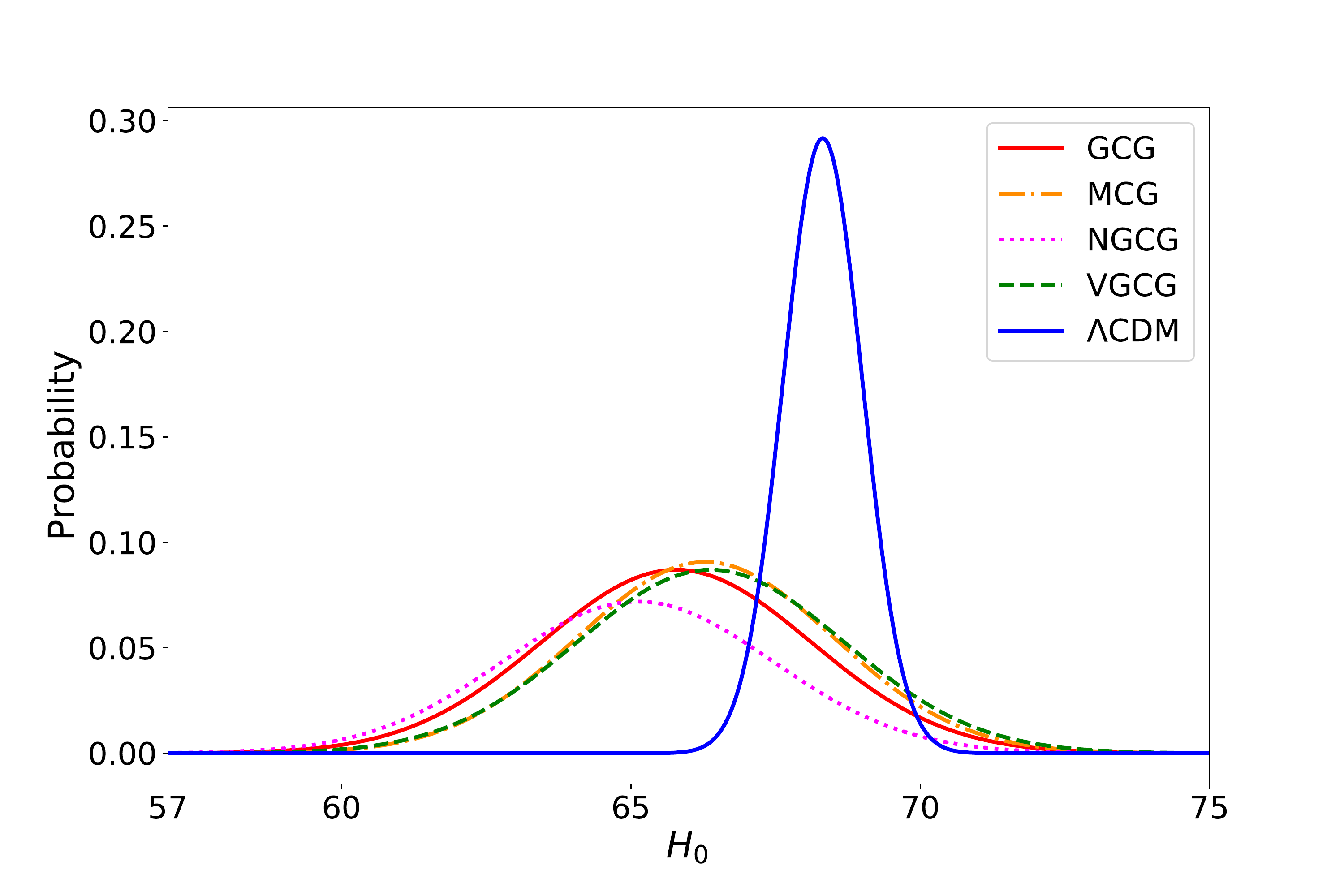}}
\subfigure{\includegraphics[width=2.5in]{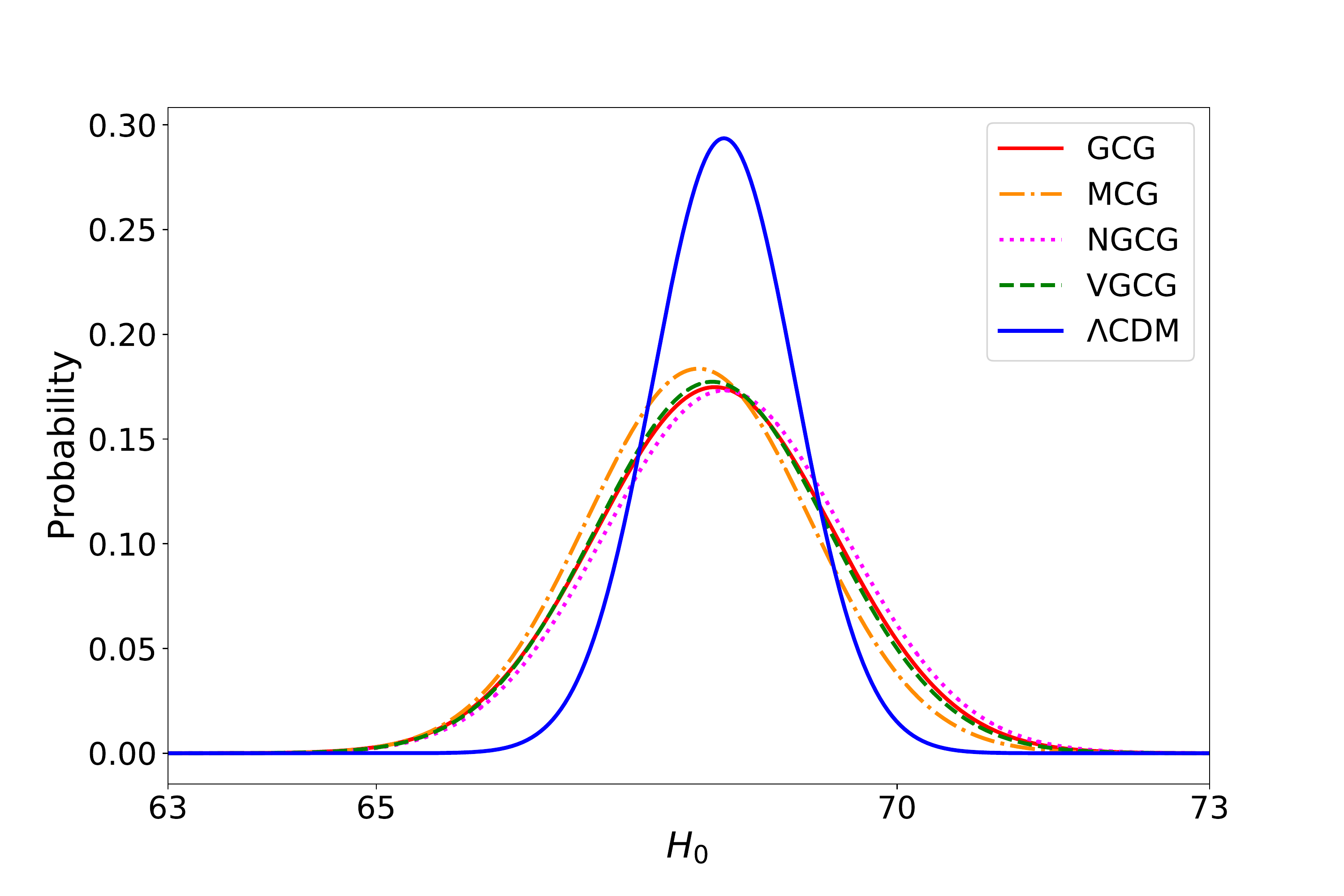}}
\caption{The posterior distributions of $H_{0}$ for the GCG, MCG, NGCG, VGCG and $\Lambda$CDM models, with the standard candles, standard rulers and combination data from the top to the bottom. We adopt the posterior distributions of $H_{0}$ from Table~\ref{tab:results}.}
\label{fig:JSD_H0}
\end{figure}
This new class of information-theoretic divergence measures based on Jensen's inequality and the Shannon entropy, called ``Jensen-Shannon Divergence", could assign the similarity between two probability distributions \cite{Lin1991JSD,JSD2012,Lian2021}. It should be mentioned that JSD is used to assess two different cosmological models by the common parameters; here, we choose the matter density $\Omega_{m}$ and the Hubble constant $H_{0}$ to distinguish the four CG models as well as the $\Lambda$CDM model.
In general, the JSD is symmetric and ranges from 0 to 1, which can be written as
\begin{equation}
D_{J S}(p \mid q)=\frac{1}{2}\left[D_{K L}(p(x) \mid s)+D_{K L}(q(x)\mid s)\right],
\end{equation}
where $s=1/2(p+q)$. $p(x)$ and $q(x)$ are two probability distributions of two different models and $D_{KL}$ denotes the Kulback-Leibler divergence (KLD), which can be expressed as
\begin{equation}
D_{K L}(p \mid q)=\int p(x) \log _{2}\left(\frac{p(x)}{q(x)}\right) d x.
\end{equation}
It is clear that a smaller value of JSD indicates that the two models are similar.
Fig.~\ref{fig:JSD_Om} and Fig.~\ref{fig:JSD_H0} display the posterior distributions of $\Omega_{m}$ and $H_{0}$. Table~\ref{tab:IC} presents the JSD values between the $\Lambda$CDM model and four nonstandard models by using different observations with respect to $\Omega_{m}$ and $H_{0}$. 
For standard candle data, the posterior distributions of $\Omega_{m}$ and $H_{0}$ in the NGCG model agree more with the $\Lambda$CDM model in terms of the JSD values, while the MCG model shows a larger distance from the $\Lambda$CDM model. In the scenario of standard ruler data, the value of JSD concerning $\Omega_{m}$ shows that the GCG model agrees more with the $\Lambda$CDM model; however, concerning $H_{0}$, all four nonstandard models are distant from the $\Lambda$CDM model, where the VGCG model is closest to the $\Lambda$CDM model.
In the case of the combination sample, for $\Omega_{m}$, the GCG model and NGCG model are more closer to the $\Lambda$CDM model due to the smaller values of JSD, while for $H_{0}$, the MCG model and VGCG model are closest to the $\Lambda$CDM model.

\subsection{Statefinder Diagnostic}
\label{sec:statefinder}
\begin{figure}
    \centering 
    \includegraphics[width=\columnwidth]{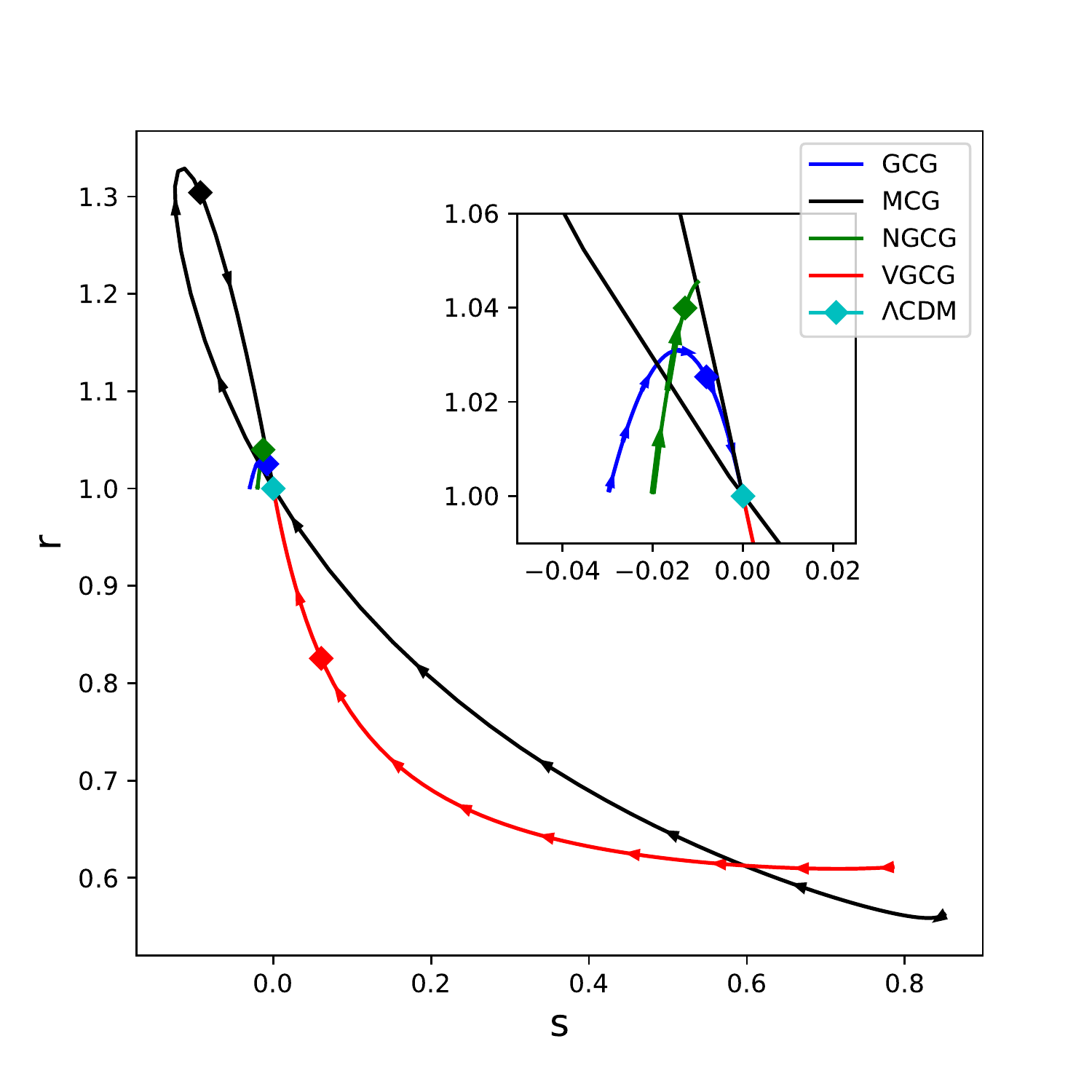}
    \caption{The evolution of the statefinder pair $(r,s)$ for different cosmological models. The cyan diamond point at $(r,s)=(1,0)$ indicates the $\Lambda$CDM model, and the other diamond point on each curve denotes the present value of the statefinder pair $(r,s)$ for the GCG, MCG, NGCG, and VGCG models. The model parameters adopted in statefinder diagnostic are from the combination of QSO[XUV], SNe Ia, QSO[AS] and BAO in Table~\ref{tab:results}.}
    \label{fig:statefinder}
\end{figure}

\begin{figure}
    \centering 
    \includegraphics[width=\columnwidth]{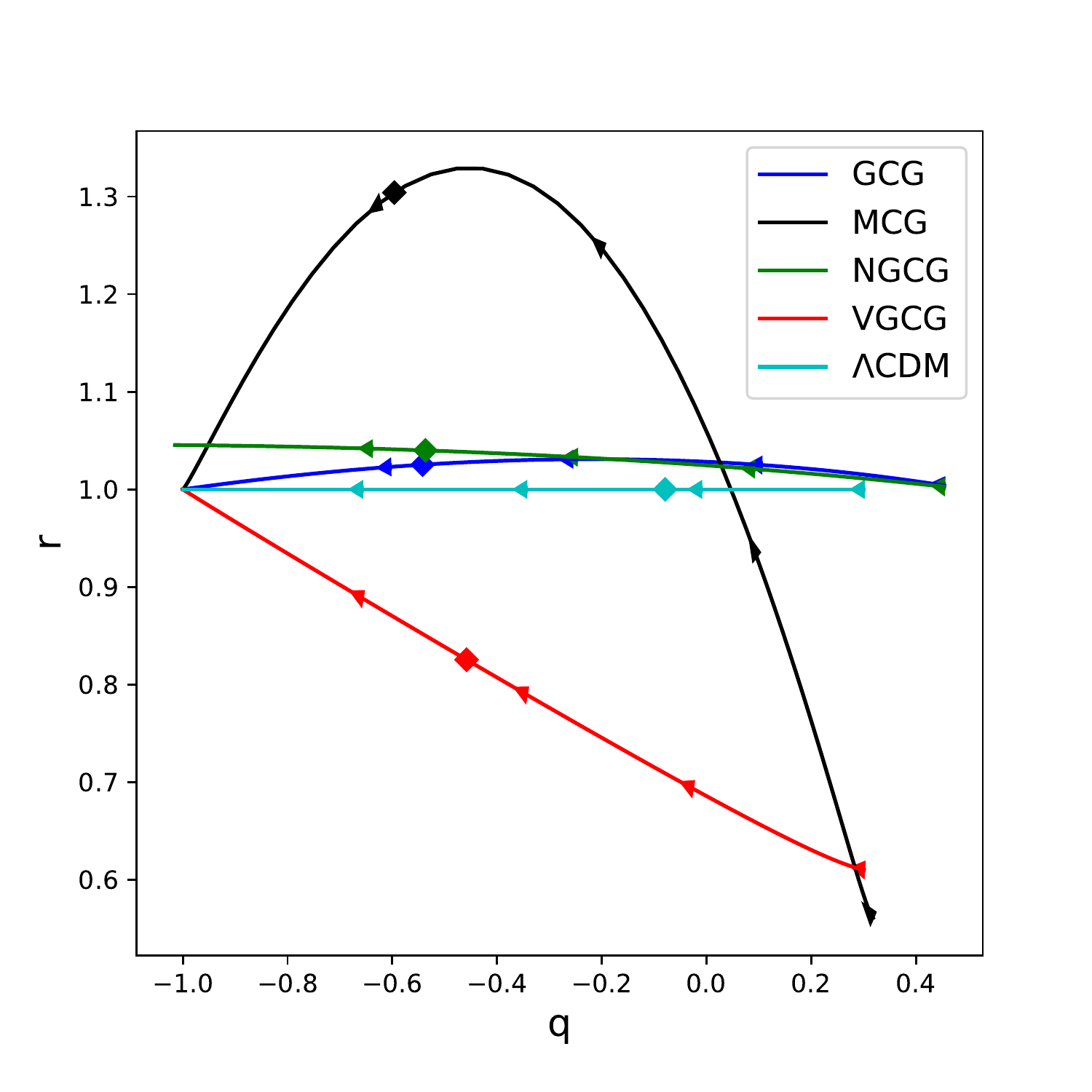}
    \caption{The same as Fig.~8, but for the evolution of the pair $(r,q)$.}
    \label{fig:rqplane}
\end{figure}

In the framework of a specific cosmological model, the Hubble parameter $H(z)$ and the deceleration parameter $q(z)$ can be expressed,
\begin{equation}
H=\frac{\dot{a}}{a}, q=-\frac{\ddot{a}}{a H^{2}}=-\frac{a \ddot{a}}{\dot{a}^{2}},
\end{equation}
where $a$ is the scale factor $a=1/1+z$. As $H(z)$ and $q(z)$ cannot effectively distinguish different cosmological models, it requires a higher order of time derivatives of $a$. To investigate more dark energy models, except for the cosmological constant model, the author of \cite{statefinder} focused on a new geometrical diagnostic pair $(r,s)$ constructed from the $a(t)$ and its third time derivatives beyond, where $r(z)$ is a natural next step beyond $H(z)$ and $q(z)$, and $s(z)$ is a linear combination of $r(z)$ and $q(z)$. This approach has been widely adopted in comparing different cosmological models \cite{lixiaolei,xutengpeng2018,dubey2021statefinder,panyu2021statefinder}.

The statefinder pair $(r,s)$ is also related to the equation of state of dark energy and its first time derivative, which can be expressed as
\begin{equation}
r=\frac{\dot a}{a H^{3}}, \quad s=\frac{r-1}{3(q-1 / 2)},
\end{equation}
For a given model, the statefinder diagnostic can be obtained by
\begin{equation}
r(z)=1-2 \frac{E^{\prime}(z)}{E(z)}(1+z)+\left[\frac{E^{\prime \prime}(z)}{E(z)}+\left(\frac{E^{\prime}(z)}{E(z)}\right)^{2}\right](1+z)^{2},
\end{equation}
and
\begin{equation}
s(z)=\frac{r(z)-1}{3(q(z)-1 / 2)},
\end{equation}
and
\begin{equation}
q(z)=\frac{E^{\prime}(z)}{E(z)}(1+z)-1.
\end{equation}
Based on the best-fit model parameters derived from the combined QSO[XUV]+SNe Ia+QSO[AS]+BAO data, we calculate the statefinder pairs $(r,s)$ for the $\Lambda$CDM model and four CG models and present the results in Fig.~\ref{fig:statefinder}. Specifically, the parameter $r$ is more effective in distinguishing different cosmological models.
It is noteworthy that although the corresponding values for the
MCG model and VGCG model significantly deviate from the $\Lambda$CDM model at the present epoch, both of them eventually converge to the standard cosmological model. On the other hand, it is obvious that in the framework of the GCG model and NGCG model, the statefinder pairs $(r,s)$ exhibit similar behaviors at present and evolve along different trajectories; however, only the GCG model ultimately converges on the point of $(r,s)=(1,0)$.

The evolutionary trajectories in the $r-q$ plane are displayed in Fig.~\ref{fig:rqplane}. Although the curves of each cosmological model originate from different points, they finally converge to the same point $(r,q)=(1,-1)$ except for the NGCG model. We clearly see that the GCG and NGCG models evolve along similar trajectories with the $\Lambda$CDM model.
In addition, we find that the GCG model and NGCG model presume values in the range $r > 1$ and $q > 0$ at early times and therefore represent as Chaplygin gas-type dark energy models. Moreover, the MCG model and VGCG model start from the regions $r < 1$ and $q > 0$ belonging to Quintessence dark energy models, while the MCG model quickly reverts back into the Chaplygin gas-type dark energy model at later times. There are notable flips from positive to negative in the value of $q$, which explains the recent phase transition of these models and proves the accelerating universe exactly.

\subsection{Model selection statistic}
\label{sec:IC}
From Sect.~\ref{sec:jsd} and Sect.~\ref{sec:statefinder}, we cannot clearly determine these four CG models with the $\Lambda$CDM model. When comparing and distinguishing different competing models, certain information criteria, such as the Akaike information criterion \cite{AIC}, the Bayes information criterion \cite{BIC}, and the deviance information criterion \cite{DIC}, would be crucial.

The AIC is based on information theory, the BIC is based on Bayesian inference, and the DIC combines heritage from both Bayesian methods and information theory \cite{DIC}. Compared with DIC, the AIC and BIC are too simple to select which model performs better by only requiring the maximum likelihood and the number of parameters within a given model rather than the likelihood throughout the parameter space \cite{2011PhRvD..84b3005C,2012ApJ...755...31C}; therefore, we apply DIC to model selection in this paper. Moreover, $\Delta$DIC is an important value which denotes the difference in values of DIC between cosmological models. In our analysis, we calculate the values of DIC and $\Delta$DIC with respect to four Chaplygin gas models and $\Lambda$CDM model for same observations. In particular, negative values of $\Delta$DIC indicates that the model fits the observations better than $\Lambda$CDM model.

The DIC was introduced by \cite{DIC} and defined as
\begin{equation}
\mathrm{DIC} \equiv D(\bar{\theta})+2 p_{D},
\label{eq:DIC_1}
\end{equation}
where $D(\theta)=-2 \ln \mathcal{L}(\theta)+C$, $p_{D}=\overline{D(\theta)}-D(\bar{\theta})$, $C$ is a `standardizing' constant depending only on the data that will vanish from any derived quantity and $D$ is the deviance of the likelihood.
The definition of DIC (i.e., Eq. (\ref{eq:DIC_1})) is motivated by the form of the AIC, replacing the maximum likelihood $\mathcal{L}_{max}$ with the mean parameter likelihood $\mathcal{L}(\bar{\theta})$ and replacing the number of parameters $k$ with the effective number of parameters $p_{D}$, which represents the number of parameters that can be usefully constrained by a particular dataset. By using the effective number of parameters, the DIC also overcomes the problem of the BIC that they do not discount parameters that are unconstrained by the data \cite{DIC}. In the DIC analysis, the favorite model is the one with the minimum DIC value.

We introduce the DIC to evaluate which model is more consistent with the observational data. As for standard candle data, it suggests that the DIC criterion advocates on the MCG model. From standard rulers and the combination sample, the VGCG model seems to be preferred by the smallest values of DIC. In addition, the GCG and NGCG model are seriously punished by the DIC. In particular, we use the model selection DIC criterion to specify which model is preferred by the currently available observations, rather than selecting the single best-fit cosmological model. As shown in the recent observational constraints on $f(T)$ gravity \cite{PhysRevD.100.083517}, the exponential $f(T)$ model presents a small deviation from $\Lambda$CDM paradigm, based on the SNe Ia Pantheon sample, Hubble constant measurements from cosmic chronometers, the CMB shift parameter and redshift space distortion measurements. Our findings demonstrate that the MCG model and VGCG model behave better than the concordance $\Lambda$CDM model. We remark here that the $\Lambda$CDM cosmological model, built on the assumptions of a cosmological constant and cold dark matter, shows a $\sim 4\sigma$ tension with the high-redshift Hubble diagram of SNe Ia, QSO and gamma-ray bursts (GRB) \cite{2019NatAs...3..272R}. Such irreconcilable tension between high-redshift QSOs and flat $\Lambda$CDM, which has been recently traced and extensively discussed \cite{2020PhRvD.102l3532Y,Lian2021} in the framework of log polynomial expansion and modified gravity theories, highlights the seriousness of the conflict with dark energy within the flat $\Lambda$CDM model. However, it is still interesting to see if future high-redshift datasets show similar tension with flat $\Lambda$CDM cosmology, given the limited sample size and current quality of the available observational data.

\section{Conclusions}

In this paper, we investigated the constraint ability of standard candles (QSO[XUV]+SNe Ia) and standard rulers (QSO[AS]+BAO) on a series of Chaplygin gas models, including the GCG model, MCG model, NGCG model and VGCG model. These Chaplygin gas models are considered as important candidate models that regard dark energy and dark matter as a unification. The first part is devoted to performing MCMC statistical analysis to confront the models with the most recent observations. The second part is dedicated to comparing the agreement between the $\Lambda$CDM model and the other four models using JSD, exploring the evolution of cosmological and cosmographical parameters with the assistance of statefinder diagnostic analysis and examining the viability of four nonstandard models by information criteria such as DIC.
Here, we summarize our main conclusions in more detail:

(i) It is intriguing that the value of $\Omega_{m}$ is noticeably larger from the standard candle data than that from other measurements. Such discrepancy is caused by the QSO X-ray and UV flux data, which favors the higher $\Omega_{m}\sim 0.5-0.6$ discussed in \cite{Risaliti2015,khadka2020_1} at high redshifts $z\sim2-5$. Therefore, the quasar data at high redshifts can cast a new light on investigating the accelerating universe. Considering the Hubble constant, it is noteworthy that the constraint results from standard candles and the combination sample suggest central values on $H_{0}$ between the value measured by the Planck CMB measurements and local $H_{0}$ measurements, possibly alleviating the tension between these measurements. In addition, it is remarkable that although we are using data based on local measurements, such as SNe Ia, which favors the local value (SH0ES's result), it does not play a role in constraining the Hubble constant caused by the marginalization of the constant term $Y_{0}$ we adopted. Hence, the QSO data from X-ray and UV flux measurement prefers the value of $H_{0}$ from the Planck 2018 results.

(ii) Most CG models include the concordance $\Lambda$CDM model as a special case corresponding to certain values of their parameters, such as the parameter $\alpha$ in the GCG model and the parameters $B$ and $\alpha$ in the MCG model. For standard ruler data, the GCG model and NGCG model are generally inconsistent with the cosmological constant case within a 68\% CL, while the MCG model and NGCG model disagree with the $\Lambda$CDM model by the combination sample at a 68\% CL. In the previous studies, they concluded that the CG model is ruled out by recent observations. In our work, considering standard candle data, the CG model is accepted in all cases. The CG model is favored in the framework of the MCG and VGCG models from standard ruler data as well as combined sample.
This is because that the inclusion of QSOs from X-ray and UV measurements and QSOs from VLBI could provide more information from the early universe. Hence, it is expected that these selected quasars could be considered additional probes in the future. 

(iii) To evaluate the similarity between $\Lambda$CDM and other CG models, we adopt the JSD in this paper. For standard candle data, the posterior distributions of $\Omega_{m}$ from four nonstandard models are distant from the $\Lambda$CDM model, while the NGCG model is in good agreement with the $\Lambda$CDM model in terms of the JSD value of $H_{0}$. For standard ruler data, the NGCG model shows a larger distance from the $\Lambda$CDM model according to the values of JSD from the posterior distribution of $\Omega_{m}$ and $H_{0}$. The posterior distributions of $\Omega_{m}$ and $H_{0}$ from the MCG model and VGCG model are in good agreement with the $\Lambda$CDM model from the combined standard candle and ruler data. Based on the best fits obtained with the combination sample, we apply the statefinder diagnostic to discriminate the dynamic behaviors of the four CG models. The GCG model and NGCG model evolve similarly to the $\Lambda$CDM model, but the NGCG model could stray from the $\Lambda$CDM model in the near future. Clearly, the MCG model and VGCG model exhibit significantly different evolutionary trajectories to the $\Lambda$CDM model; however, they approach to $\Lambda$CDM in the future.
According to the DIC criterion, VGCG model is more favored by observations; on the other hand, the GCG and NGCG models are punished by all catalogs of data. In addition, the MCG model is slightly supported by standard candles data.

In conclusion, we find that the VGCG model and MCG model could be strong candidates for investigating the accelerating universe. Moreover, $H_{0}$ tension will be alleviated with VGCG model and MCG model and these models can satisfy the combination of standard candle and standard ruler measurements with $\Delta$DIC=-6.64 and $\Delta$DIC=-0.56 compared with $\Lambda$CDM model.
In addition, it is distinctive that the CG model cannot be ruled out by high redshift observations, such as the compilation of 1598 QSO X-ray and UV measurements. Therefore, extending the cosmological analysis with high-redshift data should be critical in distinguishing between different CG models that are degenerate at low redshifts. As a result, it is promising that future precise high redshift data (i.e., gravitational wave data) will provide stronger evidence to judge whether dark energy and dark matter are unified and to understand the nature of the accelerating universe. 
There are several issues we do not consider in this paper and which remain to be addressed in the future analysis. One general concern is given
by the fact that we have considered only the 0th order cosmology and Chaplygin gas models might have instabilities at the perturbation level. Some work has also studied the behavior of the particular case of generalized Chaplygin gas models in the matter power spectrum. As worked out in detail by \cite{PhysRevD.69.123524}, the oscillations or exponential blowup of power spectrum, which are inconsistent with the observations of the 2df galaxy redshift survey, contribute to the ruling out of GCG models in 1st order cosmology (the growth of linear perturbations). Now precision data of 
redshift-space distortions (RSD) \cite{growthdata11,Arman2018,201109516}, the rms mass fluctuation $\sigma_{8}$(z) inferred from galaxy and Ly-$\alpha$ surveys \cite{growthdata12,growthdata13,Cuceu2021}, weak lensing statistics \cite{growthdata14}, baryon acoustic oscillations \cite{growthdata15,growthdata15_2}, X-ray luminous galaxy clusters \cite{growthdata16}, and Integrated Sachs-Wolfe (ISW) effect \cite{growthdata17} are gradually allowing us to determine the linear growth function that are related to perturbations. In the future analysis we will take a further step in this direction, focusing on more stringent constraints on the perturbative behaviors of a series of Chaplygin gas models.

\section*{Acknowledgements}

This work was supported by National Key R\&D Program of China No.
2017YFA0402600; the National Natural Science Foundation of China under Grants Nos. 12021003, 11690023, and 11920101003; the Strategic Priority Research Program of the Chinese Academy of Sciences, Grant No. XDB23000000;
and the Interdiscipline Research Funds of Beijing Normal University. 

\section*{DATA AVAILABILITY STATEMENTS}

The data underlying this article will be shared on reasonable request
to the corresponding author.

  

\bibliographystyle{spphys}       
\bibliography{refer}   

\end{document}